\documentclass[oldversion]{aa}

\usepackage{graphicx}
\usepackage{txfonts}

\begin{document}

\newcommand{\fe}{[\ion{Fe}{ii}]}
\newcommand{\s}{[\ion{S}{ii}]}
\newcommand{\h}{H$_2$}
\newcommand{\kms}{km\,s$^{-1}$}
\newcommand{\um}{$\mu$m}
\newcommand{\lam}{$\lambda$}

\title{Recipes for stellar jets: results of combined 
optical/infrared diagnostics
\thanks{Based on observations collected at the European Southern 
Observatory, La Silla, Chile (ESO programmes 070.C-0396(A), 070.C.-0396(B))}}

\author{Linda  Podio\inst{1} 
\and Francesca Bacciotti\inst{2} 
\and Brunella Nisini\inst{3} 
\and Jochen Eisl\"offel\inst{4} 
\and Fabrizio Massi\inst{2} 
\and  Teresa  Giannini\inst{3} 
\and Thomas P. Ray\inst{5}}
\offprints{Linda Podio, lindapod@arcetri.astro.it}

\institute{Dipartimento di Astronomia e Scienza dello Spazio, 
Universit\'a degli Studi di Firenze, Largo E. Fermi 2, I-50125 Firenze, Italy
\and INAF-Osservatorio Astrofisico di Arcetri, 
Largo E. Fermi 5, I-50125 Florence, Italy
\and INAF-Osservatorio Astronomico di Roma, 
Via di Frascati 33, I-00040 Monte Porzio Catone, Italy 
\and Th\"uringer Landessternwarte Tautenburg, 
Sternwarte 5, D-07778 Tautenburg, Germany
\and School of Cosmic Physics, Dublin Institute for Advanced Studies, 
5 Merrion Square, Dublin 2, Ireland}
 
%
\date{Received date; Accepted date}
%

\titlerunning{Stellar jets analysed with  
opt/NIR diagnostics}
\authorrunning{L. Podio et al.}

\abstract{
We examine the conditions of the plasma along
a sample of 'classical' Herbig-Haro (HH) jets located in the Orion and Vela 
star forming regions,  through combined 
optical-infrared spectral diagnostics.
Our sample includes HH 111, HH 34, HH 83, HH 73, HH 24 C/E, HH 24 J,
observed quasi-simultaneously and in the same manner 
at moderate spatial/spectral resolution. 
Once inter-calibrated, the obtained spectra cover 
a wide wavelength range from 0.6 - 2.5 $\mu$m, including
many transitions from regions of different excitation conditions. 
This allows us to probe the density and temperature stratification which 
characterises the cooling zones behind the shock fronts along the jet.
From the line ratios  we derive  the variation of
the visual extinction along the flow,  
the electron density and temperature ($n_e$ and $T_e$), the 
hydrogen ionisation fraction $x_e$, and 
the total density $n_H$ in the emission region of different lines. 
The knowledge of such parameters is essential for testing existing jet 
models and for planning follow-up high-angular resolution observations.
 
From the diagnostics of optical forbidden lines 
we find, on average, that in the examined jets, in the region 
of optical emission, 
$n_e$ varies between 50 cm$^{-3}$ and 3 10$^{3}$ cm$^{-3}$, 
$x_e$ ranges between 0.03 and 0.6, 
and the electron temperature $T_e$ 
is $\sim$1.3 10$^{4}$ K in the HH 111 and HH 34
jets, while it appears to be higher (1.8 10$^{4}$ K on average) in the
other examined jets.
The electron density and temperature derived from \fe\, lines, 
turn out to be, respectively, higher and lower in comparison to 
those determined from optical lines, 
in agreement with the fact that the \fe\, lines arise in the more
compressed gas located further from the shock front. 
An even denser component in the jets, 
with values of $n_e$ up to 10$^{6}$ cm$^{-3}$ is detected using the ratio of 
Calcium lines. 

The derived physical parameters are used to estimate the depletion onto 
dust grains of Calcium and Iron with respect to solar abundances. This   
turns out to be  quite substantial, being between  70\% and 0\% for Ca and 
$\sim$90\% for Fe.  
This leads us to suggest that the weak shocks present in the beams 
are not capable of completely destroying the ambient dust grains, 
confirming previous theoretical studies. 
We then derive  the mass flux rates, $\dot{M}_{\rm jet}$, in the flows  
using two independent methods.
Taking into account the filling factor of the emitting gas, 
$\dot{M}_{\rm jet}$ is on average 5 10$^{-8}$ M$_\odot$ yr$^{-1}$.
The associated linear momentum fluxes 
($\dot{P}_{\rm jet} = v_{jet} \dot{M}_{\rm jet}$) 
are higher than, or of the same order as, those measured in the 
coaxial molecular flows, where present, suggesting 
that the flows are jet driven.  

Finally, we discuss differences between jets in our sample.
In general, we find that higher ionisation and electron temperatures
are associated with  less dense jets.
The comparison suggests that the shock mechanism
exciting the knots along the flows has the same efficiency in
all the examined objects, and the observed differences are consistent
with the different densities, and hence cooling rates, found 
in the various flows.

}
\maketitle


\section{Introduction}

Spectral analysis is a powerful tool to investigate the nature 
and properties of astrophysical 
nebulae excited by shocks and/or energetic radiation. 
The diffuse matter in such objects produces a 
wealth of permitted and forbidden emission lines, whose excitation 
properties are reasonably well known. 
This in principle  allows us to go back from the observed line 
intensities and line  ratios to the physical 
conditions of the emitting material, which in turn provide 
essential information to test theoretical 
models and plan challenging observational programs, 
e.g. those involving interferometry.
For quite a long time line diagnostics have been used mainly to determine 
the gas electron density ($n_e$) and electron  temperature ($T_e$) 
(see, e.g. Osterbrock 1994). The spectra of line emitting regions, however, 
contain much more information and the development of a number of new 
diagnostic techniques allow us to put, directly or indirectly, 
new important constraints on the gas physics. 
The study  of stellar jets is one of the fields that has benefitted 
enormously from progress in spectral techniques. These collimated flows 
appear right from the birth of a star and indeed are believed to play a 
central role in the star formation process itself 
(see, e.g., \cite{eisloffel00}, \cite{reipurth01}, \cite{ray03}, 
\cite{bacc04}). 
In fact, jets can help clear the circumstellar environment, thus setting 
a limit to the final central mass, and may even extract most of the excess 
angular momentum from the star/disk system (\cite{bacc02}, \cite{coffey04}, 
\cite{woitas05}). Prototypical cases are the HH 34 and HH 111 jets in Orion  
(\cite{reipurth01}), also studied in this work. All the observed features in 
stellar jets, such as bright knots along the flow and giant bow shocks, 
are believed to trace shocks that develop within these highly 
supersonic flows.  
Passing through the shock front, part of the bulk kinetic energy of the flow 
is turned into thermal motions, and the gas is suddenly heated to
high temperatures (10$^5$ K or more), compressed, and partially
ionised. This situation favours the collisional excitation of upper levels
of atomic transitions. A number of radiative lines, whose type and strength
are characteristic of shock excitation, are produced. These, in turn,
radiatively cool down the gas to its original temperature. 
The whole process takes place in a limited region of space behind the shock 
front called the 'cooling zone', whose length may vary from 10$^{13}$ to 
about 10$^{15}$ cm, depending on the shock strength, the pre-shock conditions 
and the composition of the gas. The structure of radiatively cooling shocks
has been widely studied in the past by several authors
(see e.g. \cite{hartigan87}, \cite{hollenbach89}, \cite{hartigan94},
\cite{hollenbach97}, \cite{flower03}, \cite{hartigan03} and references 
therein).
Within the cooling region individual physical
quantities vary enormously, creating a stratification 
of excitation conditions that results in the production of different
kinds of lines. In ground based observations at moderate spatial 
resolution, such as those presented here, the internal
structure of the shock cooling zone remains spatially unresolved.
Nevertheless, the stratification of such regions can be investigated 
spectrally by analysing the many different lines arising from different 
layers. 

In the  optical and Near Infrared (NIR) ranges used in this work
one observes forbidden and permitted  lines of abundant atomic and
molecular species, 
such as H, He, O, S, N, Fe, C, Ca, H$_2$  (\cite{reipurth01}).  
To extract more information from these lines, several attempts 
have been made to use novel techniques by various groups. 
 B\"{o}hm et al. (1980), Brugel et al. (1981) were among the first to examine 
the potential of combining the observed line ratios using 
low spatial resolution spectra integrated over the shock cooling zone.
Hirth et al. (1997) examined  
long-slit spectra to study the spatial and kinematical properties
of the forbidden emission line regions and micro-jets of a large sample 
of T Tauri stars. 
Intensities of atomic lines  have been compared with the prediction of
shock models in, e.g., Raga \& B\"ohm (1986), Hartigan et al. (1987), 
and Hartigan et al. (1994), 
indirectly inferring quantities
such as the hydrogen ionisation fraction $x_e$ and the  intensity 
of the magnetic field in the pre-shock region. 
A new simple technique  
to measure $x_e$ and $T_e$ from the ratio of optical  lines 
was first presented by Bacciotti, Chiuderi \& Oliva (1995), and
subsequently refined in Bacciotti 
\& Eisl\"offel (1999) (hereafter BE99).  
The method, referred to as the `BE' technique, 
is based on the fact that the gas emitting forbidden lines is 
collisionally excited, but no assumption is made regarding the heating agent. 
This is clearly  an advantage if the results are used to validate
a given thermal model.
On the other hand, the method assumes that the emitting gas is 
at a single temperature, which is not true in the cooling region 
behind  a shock front. In this regard, one should consider
that the results of the BE technique are relevant to the 
region behind the shocks in which the employed lines have their peak 
emission (see the discussion and diagrams in BE99).
As we show in this paper, the stratification of temperature and densities 
present in the entire cooling zone can be traced using a larger sample of lines
appropriate for different excitation conditions.

A number of jets have been analysed with the BE technique, leading to 
the finding that jets are only partially ionised, with average $x_e$ values 
between 0.01 and 0.6. This procedure is much easier to apply than a grid 
of shock models (although it is of more limited application), and thus well 
suited for the analysis of big datasets, as those provided by high angular 
resolution observations (\cite{bacc_eisl_ray99}, \cite{bacciotti02}). 
The  main advantage in determining $x_e$, independently from the gas heating 
mechanism, is that by combining its value with the derived electron density 
one can estimate the total density, n$_H$, a fundamental parameter of the jet 
that is critical in the various models.  

Diagnostic techniques using both optical and NIR lines, on the other hand, 
provides important complementary information on jet parameters. In particular,
the combination of optical and NIR \fe\, lines gives an independent tool to 
determine T$_e$ and n$_e$, which does not rely on the choice of elemental 
abundances, in the more compressed and cooler post-shock layer where these 
lines are excited (\cite{nisini02}, \cite{pesenti03}, \cite{hartigan04}). 
In addition, the NIR H$_2$ lines provide a means to probe the molecular 
component of shock excited gas, which may give a significant contribution
to the gas cooling in low velocity, magnetized shocks 
(\cite{eislsmithdavis00}, \cite{giannini04}).

Very recently, we have investigated the advantages of a combined optical 
and NIR spectral analysis of the HH 1 jet (\cite{nisini05}) 
using a variety of diagnostic lines over a wide wavelength range 
(from 0.6 to 2.2 $\mu$m). 
The adopted procedure turned out to be extremely 
powerful for constructing detailed physical maps of the stratified media in 
the beams of stellar jets. 
In this paper,  we analyse 
a large sample of 'classical' stellar jets,  located in the Orion 
and Vela star formation regions, with the same technique.
We obtain their basic parameters, as well 
as important quantities derived from them, such as the mass flux rates.
We analyse the variations of the parameters both behind each
shock, i.e. depending on the tracers, 
and along the jet, i.e. as a function of the distance 
from the emitting source. 
We then discuss the differences/similarities among jets 
having different properties and ages. This information
can in turn be used for the selection of suitable candidates 
for high angular resolution observations.

This paper is organized as follows: we describe our observations in Sect. 2 
and briefly recall our diagnostic procedure in Sect. 3, adding a detailed 
discussion on the choice of the best set of elemental abundances, an issue 
that was not examined in our previous papers. In Sect. 4 we describe the  
results obtained for each target in the sample, and we discuss them in Sect. 5
together with the derivation of the depletion of refractory elements and the 
mass and linear momentum flux rates. Sect. 6 summarizes our findings.

\section{Observations}

We observed a sample of classical protostellar jets 
(HH 34, HH 111, HH 83, HH 73, HH 24 C/E, HH 24 J)
in the spectral range from 6015 \AA\, to 2.52 \um.
We acquired the optical spectra (6015-10320 \AA) 
at the ESO 3.6-m telescope equipped with the spectrograph EFOSC2. 
Infrared spectra were taken at the NTT with SofI using both the blue 
grism (IR-GB: 0.95-1.64 \um) and the red grism (IR-GR: 1.53-2.52 \um).
In both cases the spectra were taken aligning the slit parallel to the 
jet axis.
The position angles are: 166$^{\circ}$ for HH 34, 277$^{\circ}$ for HH 111, 
299$^{\circ}$ for HH 83, 303$^{\circ}$ for HH 73, 334$^{\circ}$ for HH 24 C/E, 
and 309$^{\circ}$ for HH 24 J.
The observational settings were chosen to obtain, as far as possible, 
homogeneous spectra in the two wavelength ranges, 
in order to be able to apply a combined optical/NIR analysis.
The observations were all carried out over a short period of time 
(7-8 and 11-12 January 2003, for the optical and IR observations, 
respectively). 
In this reagard, note that jet parameters change over timescales not shorter 
than a few years 
and probably decades in most cases. 
A 1$\arcsec$ width slit was used 
for both the optical and the IR spectra, in order to cover the 
same section of the jet and to have similar spectral resolution (R $\sim$ 600).
Moreover, the spatial scale of the two cameras is comparable 
(0$\farcs$314/pixel for EFOSC2 and 0$\farcs$288/pixel for SofI).
The integration times for the EFOSC2 observations were 1800s for 
HH 111, HH 34, HH 73 and 5400s for HH 83, HH 24 C/E, HH 24 J.
In the infrared the exposure times 
for each of the two SofI grisms are: 1800s for HH 111 and HH 34, 
1200s for HH 83 and HH 24 C/E, 
600s for HH 24 J, 800s in the IR-GB and 400s in the IR-GR for HH 73.    
In addition, telluric and spectro-photometric standards were observed in order
to correct for the atmospheric spectral response and to flux-calibrate 
the spectra respectively.
The wavelength calibrations were performed using a helium-argon lamp in the 
optical and a xenon lamp in the infrared.
The data reduction was done using standard IRAF tasks. We obtained 
for each jet three individually calibrated spectra 
in the optical, IR-GB and IR-GR ranges.
For each knot, a single spectrum from 6015 \AA\, to 2.52 \um\, was then
formed as follows:
(i) The same spatial reference was defined in each of the three spectra. 
(ii) A knot length was set by comparing the spatial profiles 
along different lines with our images acquired through [SII] 
and H$_{2}$ narrow-band filters and 
with  high angular resolution images from the literature. 
The same knot size was then adopted in all three spectral domains. 
(iii) We then extracted from each spectral image, and for each knot, 
the corresponding spectrum integrating over the size defined above.  
(iv) For each knot, the three obtained spectra were inter-calibrated measuring
the fluxes of the lines located in the overlapping spectral regions 
(the [CI]9850\AA\, line flux in the optical/IR-GB and the 
[FeII]1.64\um\, line in the IR-GB/IR-GR).

\section{Description of the combined optical/NIR spectral analysis}

In Sect. 3.1 the principles underlying our diagnostic techniques are 
briefly recalled.
In Sect. 3.2 we present a critical discussion on the adoption of a given 
set of elemental abundances, which is necessary for our analysis.

\subsection{Derivation of the physical conditions along the jet beam}

Thanks to the many lines observed over the wide spectral range covered by our 
EFOSC2/SofI spectra, and using together optical (\cite{be99}) and infrared 
(\cite{nisini02}, \cite{pesenti03}) diagnostic techniques,
we have been  able to determine the parameters describing the stratified 
medium behind each shock even if the knots along the jets are not spatially 
resolved.

Firstly, the visual extinction, $A_{\rm V}$, has to be determined
to deredden the spectra and be able to combine lines that are far apart in 
wavelength.
$A_{\rm V}$ can be found using
pairs of distant lines coming from the same upper 
level of a single ion and assuming a reddening law. 
Suitable choices in our case are 
the  \fe1.64\um/1.25\um\, or the  \fe1.64\um/1.32\um\, ratios.
We tried to calculate $A_{\rm V}$ from both these ratios and using 
different sets of Einstein coefficients
(Nussbaumer \& Storey 1988, \cite{quinet96}, \cite{smith06}).
With any of the above sets the \fe1.64\um/1.25\um\, ratios give higher 
values of $A_{\rm V}$ than those
inferred through using \fe1.64\um/1.32\um\,.
Nisini et al. (2005) pointed out that the   
\fe1.64\um/1.25\um\, ratio, compared with
the theoretical value adopting the radiative rates from \cite{nussbaumer88},
provides extinction values which are
too high, inconsistent with determinations in the optical and
with model predictions (see their Appendix B).
We find that the \fe1.64\um/1.25\um\, ratio empirically estimated by  
Smith \& Hartigan (2006) from
the spectrum of P Cygni produces even higher $A_{\rm V}$ values. 
On the other hand, Smith \& Hartigan (2006)
derive a \fe1.64\um/1.32\um\, theoretical ratio very similar to that of 
\cite{nussbaumer88}, 
which gives an $A_{\rm V}$ more consistent with other independent 
determinations (Nisini et al. 2005).
Thus we adopted the $A_{\rm V}$ value derived from the  \fe1.64\um/1.32\um\, 
ratio and the \cite{nussbaumer88} coefficients as in Nisini et al. (2005).
All the spectra were then corrected for extinction using a standard 
dereddening procedure (\cite{draine89}) and an interpolation of the extinction 
law derived by  \cite{rieke85} for the near-IR bands.

The electron density, $n_{e}$, in the region of optical emission,
was derived from the \s \lam 6716\AA/\s \lam 6731\AA\, ratio (Osterbrock 1994).
Since in some of the knots
the \s\lam\lam6716, 6731 doublet was not resolved sufficiently  to separately 
measure the line fluxes, we used 
data taken two years before by our group using EFOSC2 at higher
spectral resolution (Medves, Bacciotti \& Eisl\"{o}ffel, in prep.). 
Intervening proper motions of the jet knots can be neglected for the purposes 
of our analysis.
Then, using the BE technique (see \cite{be99}), 
that employs selected optical transitions of 
S$^{+}$, O$^{0}$, and N$^{+}$, we derived the electron
temperature, $T_{e}$, and the hydrogen ionisation fraction, $x_{e}$.
Since in determining these quantities one uses ratios between different 
species (namely [\ion{N}{ii}]$\lambda\lambda$(6548+6583)/[\ion{O}{i}]$\lambda\lambda$(6300+6363) and 
\s$\lambda\lambda$(6716+6731)/[\ion{O}{i}]$\lambda\lambda$(6300+6363)), 
the technique requires the 
adoption of a given set of elemental abundances.
This issue is examined in detail in the next Section. 
Finally, from the values inferred for  $n_{e}$ and $x_{e}$, we derived 
a gross estimate of the total hydrogen density ($n_{H}$=$n_{e}$/$x_{e}$).
Refinements to this density estimate are discussed in Sect. 5. 
Also, as mentioned in the Introduction,  
we stress that the results of the BE technique are only  relevant to
the region of the cooling zone behind the shock front where the
considered optical lines emit the most (see \cite{be99}).

The errors that affect the parameters obtained through the BE diagnostic are 
due to measurement errors of the line fluxes (which depend on the 
signal-to-noise ratio) and the uncertainty in the determination of $A_{\rm V}$ 
which is used to deredden the line fluxes.
To determine $x_{e}$ and $T_{e}$ the 
uncertainty in $n_{e}$ values also has to be taken into account.
On the other hand, since the lines used in this diagnostic are very close in 
wavelength the main source of uncertainty is the measurement error.  
In general, for the brightest knots, where the signal-to-noise is high 
(S/N $>$ 10), the errors are $<$5\% for n$_{e}$, 
$<$15\% for $x_{e}$ values and $<$10\% for $T_{e}$.

\fe\, lines give a completely independent way to determine 
the physical conditions of the gas.
From the ratios \fe1.64\um/1.53\um\, and \fe1.64\um/1.60\um\,
one can infer the electron density $n_e$ in the region of Fe emission. 
The ratios between optical and infrared \fe\, lines
can be used to find the temperature in the same region.
The relevant ratios for this goal, are those of the 1.64\um\, line with
different transitions between 8000 \AA\, and 1\um\, originating from the 
{\it a}$^4$\,P term (\cite{nisini02}, \cite{pesenti03}).
As we will show, \fe\, lines  trace the cooler and denser gas located further 
from the shock front than the zone of optical line emission,
and thus probe a region where the emitting material is more compressed 
(see \cite{hollenbach97}).

In order to check if there is an even denser component in the knots
we used the \ion{Ca}{ii}\lam8540/[\ion{Ca}{ii}]\lam7290 and 
\fe\lam7155/\lam8617 ratios.
The theoretical values of \ion{Ca}{ii} ratios for three different values of 
the temperature (T = 5000 K, T = 10000 K, T = 15000 K) were computed with a 
five-level statistical equilibrium 
code (radiative transition rates from NIST; collisional rates
from \cite{mendoza83} and \cite{chidichimo81}).
This code assumes purely collisional excitation. 
The observed ratio [\ion{Ca}{ii}]$\lambda$7290/$\lambda$7324, $\sim$1.5 in 
all the cases, 
is consistent with the ratio expected for collisional excitation.
Such a ratio is however also consistent with fluorescence pumping of the 4p
levels giving rise to the \ion{Ca}{ii} H and K UV lines, followed by 
cascade to the 3d level (\cite{hartigan04}).
Hartigan et al. (2004) estimated a neglegible fluorescence pumping rate with 
respect to the collisional rate in the \ion{Ca}{ii} excitation of the HN Tau 
jet observed at 10 AU from the source.
We assume that this is the case also for the jets in our sample, since we are 
dealing with jets located at large distance from the driving source, where no 
significant UV field is expected.
The observed values of \fe\, ratios, instead, were compared with prediction 
diagrams of \cite{bautista96}, in order to determine the electron density. 
These diagrams plot the predicted \fe\, ratios versus the
electron density for a fixed temperature 
(we considered the curve for T = 10$^{4}$ K).

Finally, to derive the gas physical conditions in the region of molecular 
emission, we use \h\, lines diagnostics. 
In particular, from the rotational diagrams we estimate the temperature and 
the total 
column density of the molecular gas (see \cite{nisini02} for more details).

\subsection{Choice of elemental abundances}

As discussed in Sect. 3.1, to estimate the ionisation fraction, x$_{e}$, and 
the temperature, T$_{e}$, one has to assume
the relative abundances of S, O, N used in the diagnostic.
As a consequence,  the derived values may depend on the adopted set of 
abundances, an issue that has to be investigated in more detail.
For example, in Nisini et al. (2005) we performed our diagnostic analysis 
adopting solar abundances from Grevesse \& Sauval (1998).
The majority of our targets, however, are located in the Orion Cloud.
Moreover, there are many lines of evidence that suggest Solar System 
abundances may not even be representative of the local ISM 4.6 Gyr ago, 
at the time the Sun formed (\cite{wilson94}). 
It follows that solar abundances may not be adequate to study the properties
of the jets in our sample.
To elucidate the situation, we test the effect of 
abundance variation on our diagnostic results.
The abundance sets we consider are:  
(i) the solar abundances determined by \cite{grevesse98} (G\&S 98),
(ii) the most recent determinations of the solar abundances by 
\cite{asplund05} (A 05),
(iii) the abundances adopted in \cite{be99} (taken
from  \cite{hartigan87}) with the aim of testing agreement
between the BE diagnostics and shock models and
(iv) the abundances determined for the interstellar gas in 
the Orion Nebula by \cite{esteban04} (E 04).
The values of elemental abundances of S, O, N in these sets  are summarized 
in Tab. \ref{tab:abundance}.

\begin{table}
\caption[]{\label{tab:abundance} Abundance  sets}

\vspace{0.5cm}
    \begin{tabular}[h]{c|ccc}
      \hline \\[-5pt]

Abundance Set & S/H         & O/H         & N/H        \\ 
              & (10$^{-5}$) & (10$^{-4}$) & (10$^{-5}$)\\[+5pt]
\hline \\[-5pt]

(i) Solar (G\&S 98)$^{a}$    & 2.14 & 6.76 & 8.32 \\
(ii) Solar (A 05)$^{b}$      & 1.38 & 4.57 & 6.03 \\
(iii) Adopted in BE99$^{c}$  & 1.6  & 6.0  & 11.0 \\
(iv) Orion (E 04)$^{d}$      & 1.66 & 4.47 & 5.37 \\ [+5pt]
\hline \\[+5pt]
      \end{tabular}

~$^{a}$\cite{grevesse98}\\
~$^{b}$\cite{asplund05}\\
~$^{c}$Bacciotti \& Eisl\"offel 1999\\
~$^{d}$\cite{esteban04}

\end{table}

\begin{figure}[!ht]
\resizebox{\hsize}{!}{\includegraphics{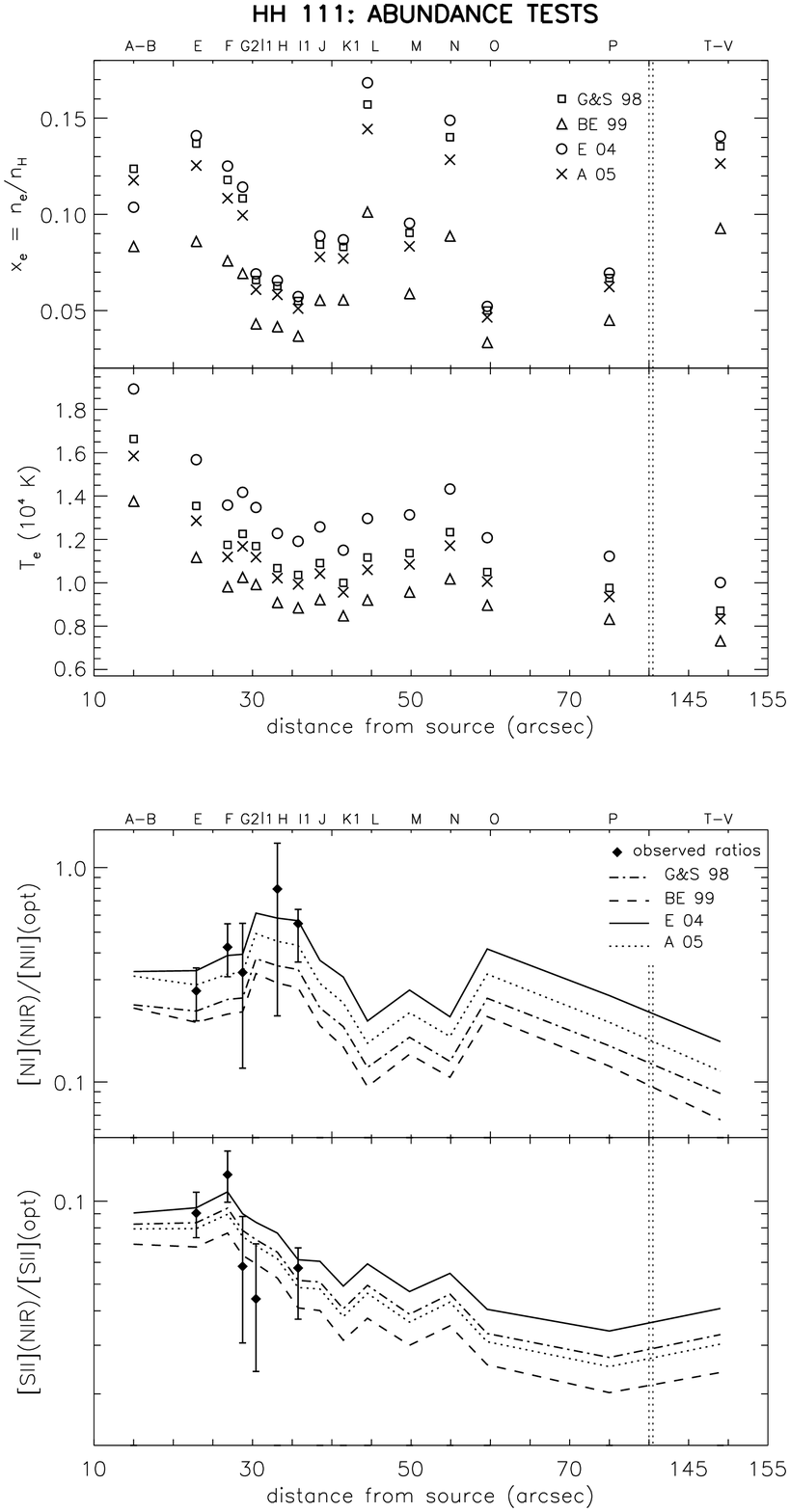}}
\caption{\label{fig:HH111_abundance} 
Variations in results using the BE technique and different adopted elemental 
abundance sets. The HH 111 jet is used for test purposes. 
{\em Upper panel}: Values of the ionisation fraction x$_{e}$ and electron 
temperature T$_{e}$ in each jet knot assuming: solar abundances from 
\cite{grevesse98} (squares) and \cite{asplund05} (crosses); 
abundances adopted in \cite{be99} (triangles); Orion abundances from 
\cite{esteban04} (circles).
{\em Lower panel}: Comparison between observed values of the 
[\ion{N}{i}](1.04$\mu$m)/[\ion{N}{ii}](6583\AA) and 
\s(1.02+1.03$\mu$m)/\s(6716+6731\AA) ratios 
and theoretical predictions for the various abundance sets.}
\end{figure}

We computed the  values of  x$_{e}$ and  T$_{e}$ 
for the  HH 111 jet assuming the above sets of abundances.
The results are shown in the upper panels of Fig. \ref{fig:HH111_abundance}.
The figure shows that with the Orion abundances of Esteban et al. (2004) we 
obtain the highest values of x$_{e}$ and T$_{e}$ from our analysis.
On the other hand, these values are within 15\% of those obtained
using the recent solar abundances determination of Asplund et al. (2005), since
the abundances of the elements involved in the diagnostics 
are quite similar in the two sets.
Moreover, the values of x$_{e}$ and  T$_{e}$ found assuming the solar 
abundances by Grevesse \& Sauval (1998) are again in good agreement with the 
previous ones.
The reason here is somewhat different: despite the Orion E 04 set and the 
solar G\&S 98 set having different abundances,
the ratios O/N and S/O used in the analysis turn out to be similar.
On the other hand, the values inferred from the abundance set adopted in BE99 
show differences of up to 40\% for  x$_{e}$ and up to 25\% for  T$_{e}$.
We also tested the results obtained with the different sets using the observed 
ratios [\ion{N}{i}](1.04$\mu$m)/[\ion{N}{ii}](6583\AA) and 
\s(1.02+1.03$\mu$m)/\s(6716+6731\AA), that do not depend on abundances,
but do depend on the derived  x$_{e}$ and  T$_{e}$.
Comparing observed and theoretical values (see the lower panels of Fig. 
\ref{fig:HH111_abundance}) we find that the abundance sets (i), (ii) and (iv) 
are all consistent with the observations, 
while the abundances adopted in \cite{be99}, that yield the lowest theoretical
values for these ratios, deviate more.

Following the above results, we chose to adopt the abundance set by 
Esteban et al. (2004) for the jets in our sample located in the Orion cloud.
For HH 73, located in Vela, we tentatively assume the same,
since no abundance determination exists for Vela.

\section{Diagnostic results: basic physical parameters of the examined jets}

In this Section we describe our determination of basic physical parameters, 
i.e., density, temperature and ionisation fraction, in each jet knot,
obtained from the application of the combined optical/NIR technique to our 
sample.
The best data were obtained for the HH 111 jet, allowing us to get
very accurate values in this case.
Parameters for HH 34 also appear to be very reliable.
HH 111 and  HH 34 are very well known HH objects and they
have been investigated previously both in  
the optical wavelength range
(\cite{eisloffel92}, Morse et al. 1993, Noriega-Crespo et al. 1993, 
Hartigan et al. 1994, Eisl\"offel \& Mundt 1997, 
Reipurth et al. 1997, \cite{be99}, Hartigan et al. 2001, Reipurth et al. 2002, 
Raga et al. 2002) and in the near-infrared (Stapelfeldt et al. 1991, 
Stanke et al. 1998, Davis et al. 2001, Nisini et al. 2002).
Our purpose here is to combine the information coming from the different 
tracers emitting in the two wavelength ranges, in order to obtain a detailed 
physical map of the stratified medium in each unresolved cooling zone along 
these outflows.
The fainter objects HH 83, HH 73, and HH 24, instead, 
were either not seen in the IR (HH 73) or were
visible only in one or two knots (HH 83 and HH 24). 
Thus for these jets we only apply the optical
diagnostic to obtaine a description of the physical conditions of the gas 
emitting in this range.
Note that, although it is not possible to estimate the extinction when the 
\fe\, lines are not detected, we
can still use the optical diagnostic since the lines involved in the BE 
technique are very near in wavelength.
A summary of our results is given in Tab. \ref{tab:physical_parameters}, 
but note that the derived physical parameters are averages for the brightest 
knots.

\subsection{HH 111}

\begin{figure}[!ht]
\resizebox{\hsize}{!}{\includegraphics{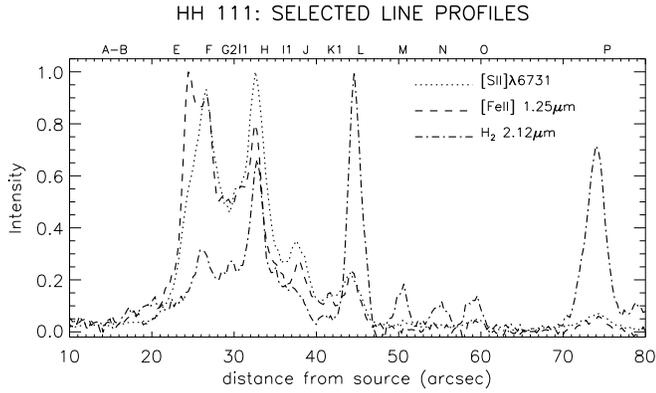}}
\caption{\label{fig:HH111_profiles} Spatial profiles of the \s,
\fe\,,  and  H$_2$ 2.12 \um\, lines along the HH 111 jet.
The zero point of the spatial scale is the source HH 111 IRS 
(RA(2000): 05 51 46.1, Dec(2000): +02 48 30).}
\end{figure}

\begin{figure}[!ht]
\resizebox{\hsize}{!}{\includegraphics{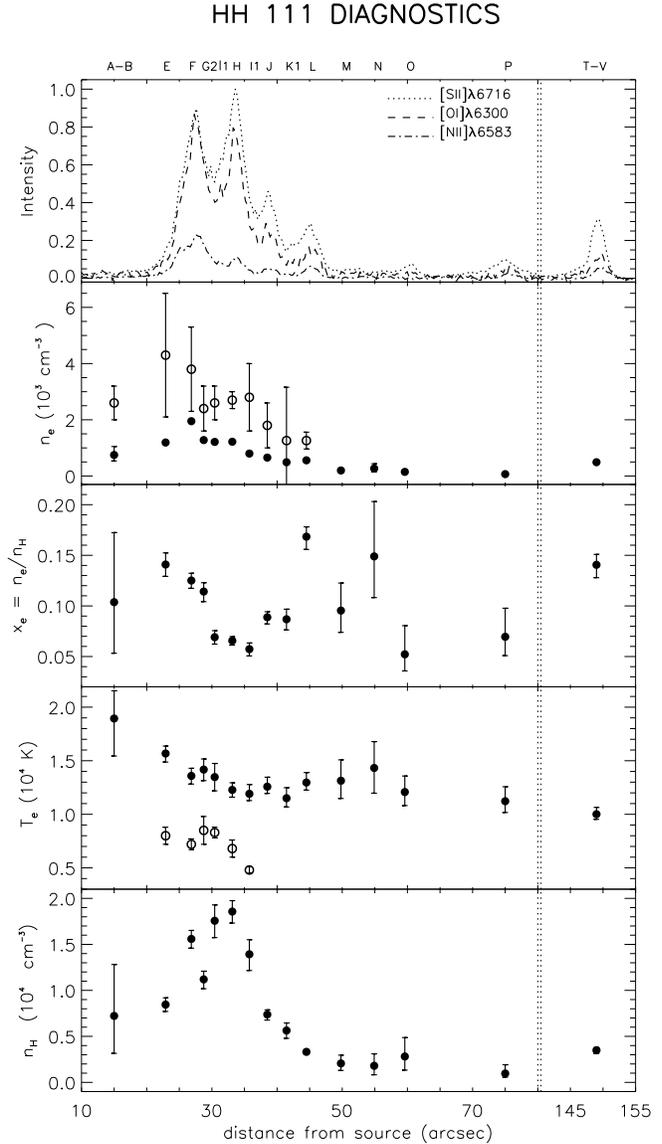}}
\caption{\label{fig:HH111_diagnostics} 
Variation of the derived physical parameters  
along the HH 111 jet. 
From top to bottom panel: 
intensity profiles of the optical lines, the electron density, n$_e$, 
in units of  10$^{3}$ cm$^{-3}$, the ionisation fraction, x$_e$, 
the temperature, T$_e$, in units of 10$^{4}$ K
and the total density, n$_H$, in units of 10$^{4}$ cm$^{-3}$.
The open circles are the values derived from the
\fe\, lines, while the filled circles are parameters inferred from the
optical S$^{+}$ ,N$^{+}$ and O$^{0}$ lines using the BE technique.
Note that the zero point of the spatial scale is the driving source 
HH 111 IRS (see Fig.\ref{fig:HH111_profiles}).
}
\end{figure}

\begin{figure}[!ht]
\resizebox{\hsize}{!}{\includegraphics{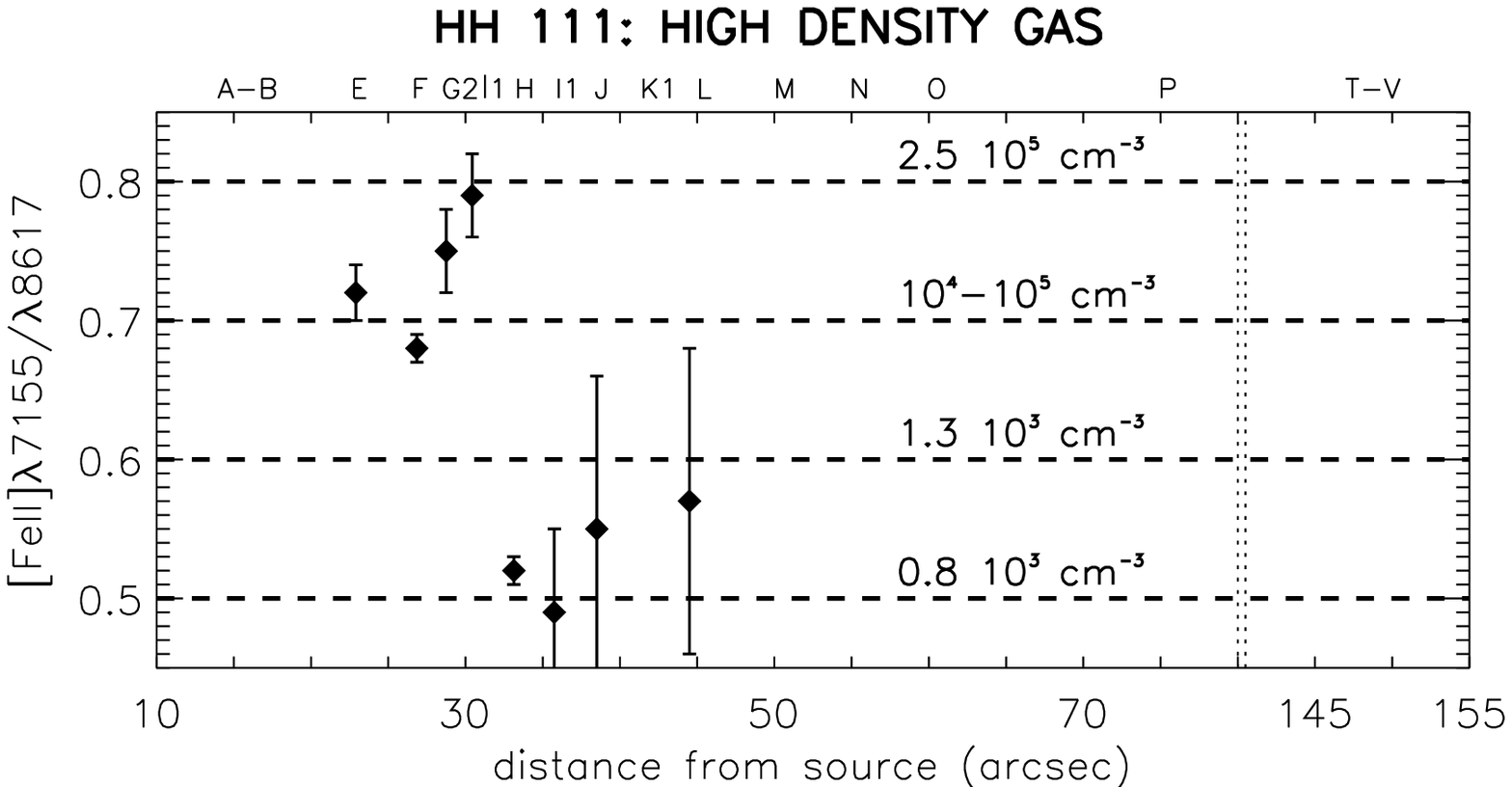}}
\caption{\label{fig:HH111_highdensity_Fe} Comparison between the observed
\fe\lam 7155/\fe\lam 8617 ratios (diamonds) and the theoretical curves 
(dashed lines) computed by \cite{bautista96} for different values of the 
electron density, n$_e$, and a fixed temperature of 10$^{4}$ K. This ratio
traces a dense component in the jet beam (n$_e$ up to 2.5 10$^{5}$ cm$^{-3}$ 
in the knots close to the source).
Note that the zero point of the spatial scale is the driving source HH 111 IRS 
(see Fig.\ref{fig:HH111_profiles}).
}
\end{figure}

\begin{figure}[!ht]
\resizebox{\hsize}{!}{\includegraphics{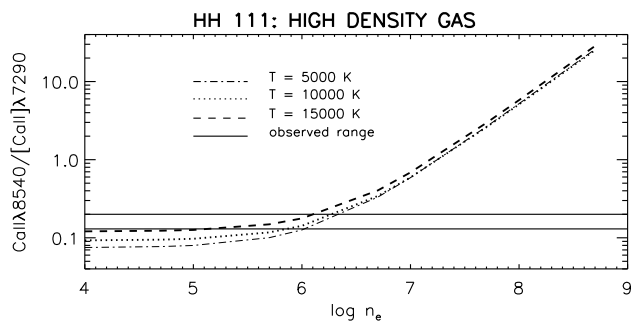}}
\caption{\label{fig:HH111_highdensity_Ca} Comparison between observed and 
predicted \ion{Ca}{ii}\lam 8540/[\ion{Ca}{ii}]\lam 7290 ratios. 
The solid lines indicate the 
range of variation of the observed ratio along the jet. The same ratio is 
computed for three different 
temperatures with a five-level statistical equilibrium 
code (dotted, dashed and dash-dotted curves). From this diagnostic,
values of the electron density of
$\sim$ 6 10$^{5}$ - 2 10$^{6}$ cm$^{-3}$ are found.
}
\end{figure}

One of the best known stellar jets is HH 111. 
This outflow, located in the L1617 cloud in Orion (D = 460 pc), is powered 
by the young  
star IRAS 05491+0247 and extends over several parsecs (\cite{reipurth97}).
The exciting source is deeply embedded in the parental molecular cloud core
(in our spectra the source is detected only in the infrared K-band) and the 
base of the jet  
is not visible at optical wavelengths. 
The main visible jet emerges from the cloud only 15$''$ from the source 
and is blue-shifted.
Adopting the nomenclature of Reipurth et al. (1997),
this lobe consists of a long chain of knots, with a bow-like morphology, 
observable in the optical 
from 15$''$ (knot A-B) to 80$''$ (knot P) and terminates with a bow shock 
(knot T-V) 
at a distance of 150$''$ from the driving source.
The red-shifted lobe is almost completely obscured, with only knot ZL visible 
in the optical,
while knots ZO and ZV (symmetric to knots O and T-V in the blue lobe) 
have been observed only in the near-infrared.

Several authors have investigated the physical properties of this flow 
in the past.
Morse et al. (1993) measured the electron density along the jet from the
[SII] line ratio and presented a complete list of the lines emitted
in the optical range (see also \cite{noriega93}). 
Hartigan et al. (1994) estimated a value of the ionisation fraction
averaged along the entire flow comparing the observed line ratios
with the prediction of shock models, and derived an estimate of the
average mass loss rate. Internal gas kinematics has been
investigated by Hartigan et al. (2001) and Raga et al. (2002). In particular, 
in the first paper proper motions of the knots were determined, 
from which shock velocities of about 40 \kms were derived, in
agreement with the type of line observed. 
In the infrared wavelength range, Davis et al. (2001) studied the
kinematics of the gas emitting in H$_2$, while 
 an analysis of H$_2$ and \fe\,  lines along the 
jet has been presented in Nisini et al. (2002).  

In this paper we apply our combined optical/NIR analysis to 
the knots in the blue lobe from 15$''$ outwards, 
where both optical and infrared lines are observed 
(except for knot T-V that is out of the frame in the IR spectra).
In Fig. \ref{fig:HH111_profiles} the spatial normalized distribution of the 
optical \s\, line is compared with the corresponding \fe\, and 
H$_2$ 2.12 \um\, distributions.
\s\, and \fe\, lines have similar shape and brightness, while
the H$_2$ line is fainter and has a different spatial distribution.
In particular the emission of \s\, and \fe\, lines is maximum at knot E with a 
flux of $\sim$ 4 10$^{-14}$ erg s$^{-1}$ cm$^{-2}$, while the maximum 
\h\, emission is reached in
knot P ($\sim$ 1.4 10$^{-14}$ erg s$^{-1}$ cm$^{-2}$).
Moreover, the \s\, and \fe\, line intensities decrease with distance from
the source, while \h\, emission is stronger in the outer bows (L and P).
Such  behavior, also noted by Nisini et al. (2002), 
can be explained by the fact that \fe\, and \s\ emission is expected to be
excited from dissociative J-shocks
($v_s$ $>$ 30-40 km s$^{-1}$) at the apex of the bows, 
while the \h\, emission arises 
from the bow wings where the transverse component
of the velocity gives rise to slower C-shocks ($v_s$ $ < $ 30 km s$^{-1}$) 
that prevent \h\, dissociation (\cite{davis96}, \cite{hollenbach97}, 
\cite{davis99}).
Observational confirmations for such an H$_2$ distribution is seen 
for example in the HH 211 flow (\cite{gueth99}).
In this scenario the fact that the (\fe\,,\s\,)/\h\, observed ratio
decreases with distance from the source would be
consistent with the suggestion that in the outer bows there is a
prevalence of non-dissociative slow C-shocks, since the ambient density 
becomes higher or comparable to the jet density (\cite{hollenbach89}).

In order to apply the diagnostics to line ratios, 
we first determined the visual extinction towards all the knots.
As expected, the extinction is quite low outside the dense core around 
the source,
being $A_{\rm V}$ $\sim$ 2 mag for the innermost visible knot (A-B), and 
decreasing to $A_{\rm V}$ $\sim$ 0 mag for the farthest knots (from I1 to T-V).
These values are much smaller than those derived in Gredel et al. (1993) and 
in Nisini et al. (2002), 
who adopted the 1.64/1.25\um\, ratio for the $A_{\rm V}$ determination, 
and smaller but more similar, at least for the inner knots, to 
the values derived from the Balmer decrement (\cite{morse93}, 
\cite{noriega93}). 
As explained in  Sect. 3.1 and, more in detail, in the appendix of 
Nisini et al. (2005), here we assume
that the 1.64/1.32\um\, ratio yields a more reliable estimate of the
reddening than the 1.64/1.25\um\, ratio when the \cite{nussbaumer88} 
Einstein coefficients are considered.

After correcting line fluxes for reddening, we derive
the values of the physical parameters, n$_{e}$,  x$_{e}$, T$_{e}$, and n$_{H}$,
separately for the different lines considered, 
for each knot along the jet, with the procedures described in Sect. 3.1.
The results are shown in Fig. \ref{fig:HH111_diagnostics}.
The electron density n$_e$ 
from  \s\, lines (filled circles) increases 
from $\sim$ 7.5 10$^{2}$ cm$^{-3}$ 
in the first knot, A-B, to $\sim$ 10$^{3}$ cm$^{-3}$ in the brightest
optical knots, reaching a maximum value of $\sim$ 2 10$^{3}$ cm$^{-3}$ 
in F. 
Then it slowly decreases with distance from the source down to 
n$_e$ $\sim$ 0.7 10$^{2}$ cm$^{-3}$ at knot P, and finally presents another 
local maximum in
knot T-V (n$_e$ $\sim$ 5  10$^{2}$ cm$^{-3}$).
These values agree with the average electron density of 900 cm$^{-3}$ 
found by Hartigan et al. (1994), while
they are of a factor of two lower with respect to the values 
found by Morse et al. (1993).
The electron temperature, T$_e$, derived from the optical lines is 
similar in all the various knots, i.e. cooling regions along the jet, 
being on average 
$\sim$ 1.3 10$^{4}$ K, if exception is made for the first
noisy  point at the base of the jet (T$_e$ $\sim$ 1.9 10$^{4}$ K).

The values of the hydrogen ionisation fraction x$_e$, in each knot,
and averaged over the region of the cooling zones 
where the considered optical lines emit the most, 
vary from a minimum of 0.05 to 0.17.
In our determination,  x$_e$ decreases steadily from knot E 
to knot K1, like the electron density, but it 
increases again at knot L to a value of 0.17. In the subsequent faint
region of the beam  the 
ionisation fraction appears to decrease again, although the accuracy
of the derivation is much lower here. Finally another local maximum is 
retrieved at the isolated knot T-V very far from the source, 
where x$_e$ is 0.14. 
The lower limit found for the jets brightest knots is 
similar to the one (x$_e$ = 0.052) derived  by Hartigan et
al. (1994) by comparing the  observed   
[\ion{N}{ii}]/[\ion{O}{i}] ratio, 
which is sensitive to x$_e$, to the predictions of
low velocity shock models. In other knots, however, we find higher values 
that are difficult to justify, as a shock with effective velocity of 
30-40 km s$^{-1}$
hardly produces an ionisation fraction greater than a few percent 
(see Fig.1 in \cite{hartigan94}). 
The contradiction can be reconciled if one considers that the shock 
fronts in the jet beam advance in the wake of the previous shock, and thus  
move in a medium that has been already pre-ionised. We note that a substantial 
level of hydrogen ionisation can persist between two consecutive shocks
because the recombination time is  slow in the rarefied jet gas. 
It can be shown that with the typical electron densities and bulk velocities 
of the jet material, the recombination  
time is of the same order of the crossing time of the entire bright section 
of the jet (see \cite{be99}).
On the other hand, this explanation may not apply to the relatively high 
ionisation of  knots L and T-V, located at the end of the two sections of the 
jet. 
In this case a higher ionisation may be produced because the shock
propagates into a medium of lower pre-shock density (see \cite{hartigan94}).

From n$_e$ and x$_e$ we obtain a gross value of the
total density of 10$^3$ - 1.9 10$^4$ cm$^{-3}$. This is maximum
for the brightest knots, as expected (the emission in collisionally 
excited lines is proportional to  n$_e$$\cdot$n$_H$).

The empty circles in the diagrams of Fig. \ref{fig:HH111_diagnostics} 
are the values of n$_e$ and T$_e$ inferred from the \fe\, lines.
From these lines  we find a higher electron density, 
n$_e$(\fe) $\sim$ 1.3 - 4.3 10$^3$ cm$^{-3}$, and 
a lower temperature, T$_{e}$(\fe) $\sim$ 4800 - 8500 K 
than inferred from the optical line ratios. 
Our result confirms that the observed \fe\, emission typically comes from 
regions of the post-shock zone that are more distant from the front than the 
optical lines, where the gas is cooler and more compressed.
Such a situation is described in detail in Nisini et al. (2005).
Note also that our values of n$_e$ are lower than those found in 
Nisini et al. (2002) 
(n$_e$ $\sim$ 0.5 - 2 10$^4$ cm$^{-3}$). This is due to the fact that 
here we use the \fe1.64\um/1.32\um\,
ratio to determine extinction, leading us to infer lower values of A$_V$.

Even higher density components can be traced through the 
\fe\lam 7155/\fe\lam 8617 and \ion{Ca}{ii}\lam 8540/[\ion{Ca}{ii}]\lam 7290 
ratios, as illustrated in Sect. 3.1 (Figs.
\ref{fig:HH111_highdensity_Fe} and \ref{fig:HH111_highdensity_Ca}).
From the \fe\, ratio we infer values of n$_e$ 
between $\sim$ 10$^{4}$ cm$^{-3}$ and $\sim$ 2.5 10$^{5}$ cm$^{-3}$ in the 
closest knots E, F, G2, and G1, that subsequently decreases to  
$\sim$ 10$^{3}$ cm$^{-3}$ in the outer knots H, I1, J, and L. 
From the [CaII] ratio we find n$_e$ $\sim$ 6 10$^{5}$ - 2 10$^{6}$ cm$^{-3}$ 
in knots E, F, G2, G1, H, J, while in the other knots no estimate was 
possible, because the signal-to-noise is too low.
The values of n$_e$ inferred from \fe\lam 7155/\fe\lam 8617 and  
\ion{Ca}{ii}\lam 8540/[\ion{Ca}{ii}]\lam 7290 
ratios demonstrate that a component of material even denser than that traced 
by \fe\, lines at 
1.64, 1.60, and 1.53\um\, is present in the knots along the jet.
If we consider for this higher density component the same ionisation 
fraction found from the optical lines,
one obtains that n$_H$ may be as high as 
 $\sim$ 10$^{7}$ cm$^{-3}$ in this layer of the post-shocked regions.
Similar high densities have been found in the HH 1 jet (\cite{nisini05}) and
at the base of a few jets from optical T Tauri stars (\cite{hartigan04}).

Using the \h\, lines flux we derived the conditions of the molecular gas in 
the jet.
In the \h\, emitting regions the temperature 
varies between 2000 and 3000 K, and the column density, N$_{H_2}$, between 
5 10$^{16}$ and
5 10$^{17}$ cm$^{-2}$. From the latter value divided by the knot size 
transverse to the line of sight, 
we derive an approximate \h\, density of $\sim$ 10$^{2}$ cm$^{-3}$. 
This is only a lower limit because 
it is supposed that the \h\, lines are  excited in a thin layer 
of molecular material located in the lateral wings of the bow 
shocks (see above). Thus the dimension of
the emitting region is probably smaller than the knot size.
Such a picture could  be confirmed by high angular resolution 
images. Nevertheless, the \h\, lines trace colder and less dense gas 
and for this reason the \h\, emission
is maximum in the outer knots L and P (knot T-V is out of the frame in the 
NIR range) 
as shown by the spatial profiles of Fig. \ref{fig:HH111_profiles}.

\subsection{HH 34}

\begin{figure}[!ht]
\resizebox{\hsize}{!}{\includegraphics{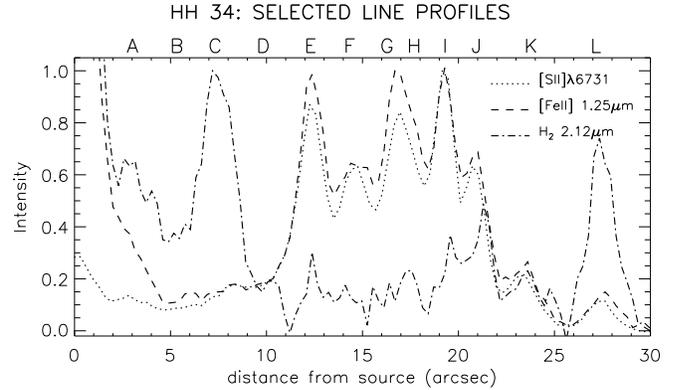}}
\caption{\label{fig:HH34_profiles} Same as Fig. \ref{fig:HH111_profiles}, 
but for the HH 34 jet.
The zero point of the spatial scale is the source HH 34 IRS 
(RA(2000): 05 35 29.8, Dec(2000): -06 26 57).
}
\end{figure}

\begin{figure}[!ht]
\resizebox{\hsize}{!}{\includegraphics{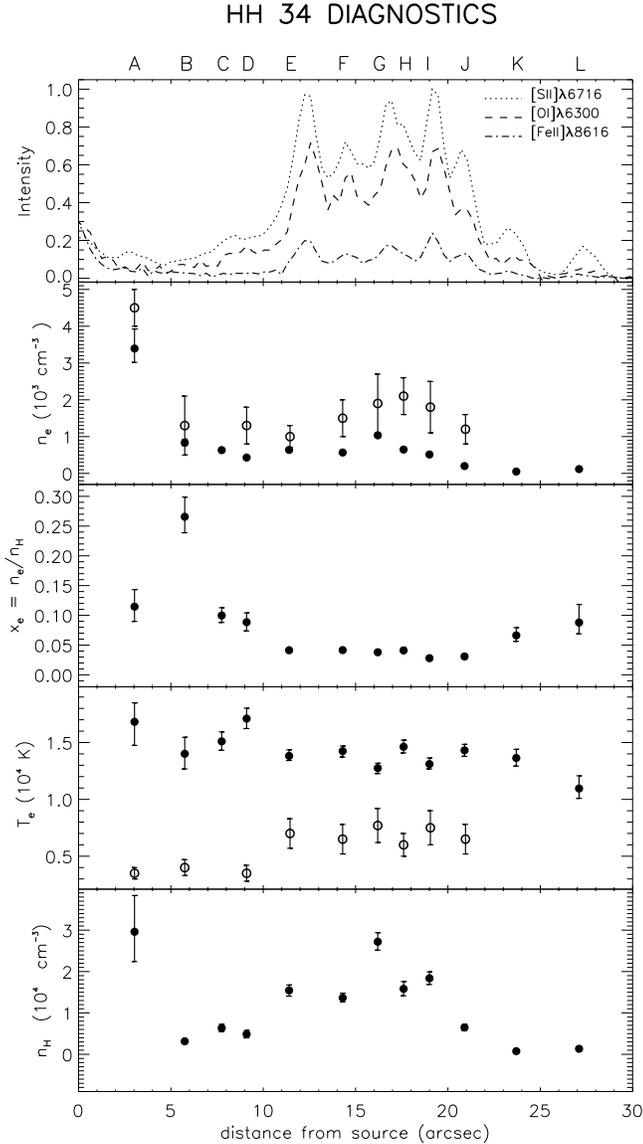}}
\caption{\label{fig:HH34_diagnostics} 
Same as Fig. \ref{fig:HH111_diagnostics}, but for the
HH 34 jet. Open circles are the values derived from \fe\, line ratios, 
while filled circles indicate the values determined using the BE technique. 
Note that the zero point of the spatial scale is the driving source HH 34 IRS 
(see Fig.\ref{fig:HH34_profiles}).
}
\end{figure}

\begin{figure}[!ht]
\resizebox{\hsize}{!}{\includegraphics{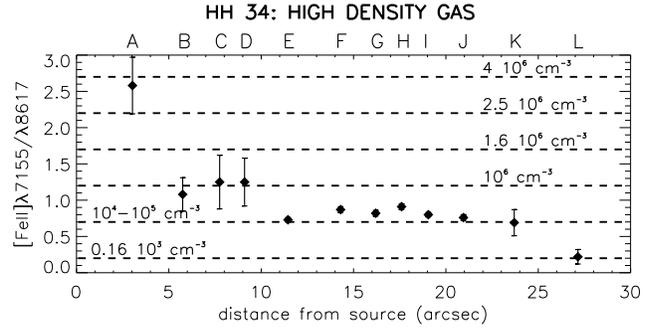}}
\caption{\label{fig:HH34_highdensity_Fe} 
Same as Fig. \ref{fig:HH111_highdensity_Fe}, but
for the HH 34 jet. Values of n$_e$ up to 4 10$^{6}$ cm$^{-3}$are found. 
Note that the zero point of the spatial scale is the driving source HH 34 IRS 
(see Fig.\ref{fig:HH34_profiles}).
}
\end{figure}

\begin{figure}[!ht]
\resizebox{\hsize}{!}{\includegraphics{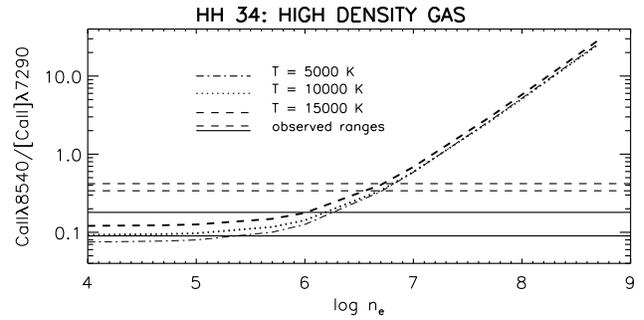}}
\caption{\label{fig:HH34_highdensity_Ca} 
Same as Fig. \ref{fig:HH111_highdensity_Ca}, but
for the HH 34 jet. The horizontal dashed and solid lines indicate 
the observed range in the knots close to the source (A, B, C, D) 
and in the outer knots (from E to L), respectively.
Electron densities up to 6 10$^{6}$ cm$^{-3}$ are found in the knots closer 
to the source.
}
\end{figure}

The spectacular HH 34 jet is located in the  L1641 cloud in Orion, and
is one of the best studied examples of stellar jets. 
It consists of  a parsec-scale flow (\cite{bally94}) the first 30$''$ of which
on the blue-shifted side is a  well-aligned chain 
of knots emitting in both optical and NIR
lines, placed south of the central source, HH 34 IRS (\cite{eisloffel92}).
HST images of the flow  reveal clearly that each knot 
has the morphology of a mini-bow shock (or 'working surface', 
see, e.g., \cite{ray96}, \cite{reipurth02}).
Two larger clumpy bow-shocks (HH 34N and HH 34S) 
symmetrically placed at a distance of $\sim$ 100$''$ 
from the source are also visible in the optical 
(see, e.g., \cite{morse92}), while the counter-jet
is not detected at optical or infrared wavelengths.
Proper motions of the knots have been measured from the ground by 
Eisl\"offel \& Mundt (1992)
and \cite{devine97}, while Reipurth et al. (2002) 
made similar measurements with HST deriving typical
shock velocities for the working surfaces along the beam of only 
20 km s$^{-1}$.  
Investigations of the physical properties of the jet gas through spectral
diagnostic, limited to optical lines, have been carried out  by
Hartigan et al. (1994) and  \cite{be99}.
In the infrared band, the jet has been investigated by
Stapelfeldt et al. (1991), who imaged the jet in \fe\, lines at moderate
and high angular resolution.
The only structure previously reported north of the star
in the H$_{2}$ 2.12 \um\ line, using a narrow-band filter, 
is an extended arc just south of the optical HH 34N (\cite{stanke98}).   

We apply our combined diagnostics to the innermost 30$''$ of the jet beam
where bright optical/NIR lines are seen.
We define 12 knots along the jet, following the optical nomenclature of 
Eisl\"offel \& Mundt (1992). 
In Fig. \ref{fig:HH34_profiles} we present the emission profiles in the 
optical \s\,  line, 
the infrared \fe\, and H$_{2}$ lines, each normalized to its intensity peak.
The atomic lines are similar both in spatial profile and brightness.
The lines intensity is maximum in knot E 
(flux $\sim$ 2 10$^{-14}$ erg s$^{-1}$ cm$^{-2}$)
and then decreases with distance from the source.
The H$_{2}$ line, on the contrary, is fainter and has a different distribution 
with two local maxima in knot C 
(flux $\sim$ 2.5 10$^{-15}$ erg s$^{-1}$ cm$^{-2}$)
and knot E (flux $\sim$ 1.3 10$^{-15}$ erg s$^{-1}$ cm$^{-2}$).
As mentioned for HH 111, this behavior is due to the 
fact that  \s\, and \fe\, lines are excited by the J-shock at the apex
of the bows, while \h\, lines are prevalent in slow, non-dissociative shocks, 
so the relative  spatial distribution of the lines
depend on the prevalence of C or J shocks along the jet.
We note also faint \h\,  emission between 20$''$ and 
31$''$, and between 69$''$ and 80$''$ north of the star.
The latter region corresponds to the arc detected by Stanke et al. (1998).
This may indicate the presence of a counterjet that is very extincted with 
respect to the blue-shifted lobe.

The visual extinction (see Sect. 3.1) is more or less 
constant in the blue lobe with a value $A_{\rm V}$ = 1.3 mag in the first 
30$''$ from the source, 
except for the nearest knot (A), for which we derive 
$A_{\rm V}$ $\sim$ 7.1 mag 
in agreement with the high extinction value found at the source by 
Reipurth et al. (1986).

The physical parameters obtained applying the BE technique to the optical 
lines emitted by each cooling zone traced by the knots 
are shown in Fig. \ref{fig:HH34_diagnostics}.
The inferred values of the electron density averaged over each cooling zone, 
n$_{e}$, decrease from a maximum of
3.4  10$^3$ cm$^{-3}$ in the first knot (A) down to 4.3 10$^2$ cm$^{-3}$ in 
knot D.
Then n$_{e}$ increases again up to a local maximum of 
$\sim$ 10$^3$ cm$^{-3}$ in knot G, and
from this point on n$_{e}$ starts to decrease again down to  
$\sim$ 0.5 - 1 10$^2$ cm$^{-3}$ in the outer knots K and L.
The ionisation fraction,  x$_{e}$, is generally quite low in all the knots.
It increases from 0.11 in the first knot, A, up to 0.27
in knot B; then decreases gently all along the bright beam down to 0.03 
in knots I and J.
From there it shows isolated higher values up to 0.08 in knots K and L.
The temperature, T$_{e}$, derived from the BE technique, turns out to
be on average $\sim$ 1.4 10$^4$ K. 
From x$_{e}$ and n$_{e}$ we obtain an estimate of the total hydrogen density, 
that shows 
a trend similar to the electron density. The maximum n$_{H}$ value is found
in knot A (n$_{H}$ $\sim$ 3 10$^4$ cm$^{-3}$), then increases from 
3 10$^3$ cm$^{-3}$ in knot B up to 2.7 10$^4$ cm$^{-3}$ in knot G. 
Subsequently  it decreases again down to 
$\sim$  10$^3$ cm$^{-3}$ in knots K and L.
  
As noticed for the HH 111 jet in Sect. 3.2, the inferred x$_{e}$ and T$_{e}$ 
values are higher with respect to those 
obtained for the same object in \cite{be99}, due to the differences between 
the adopted abundances sets.
Nevertheless, the trends are the same.  
The values of n$_{e}$ and x$_{e}$
found for HH 34 agree with the electron density derived by Morse et
al. (1993), and with
the average value of the ionisation (x$_e$ = 0.026) estimated  in
Hartigan et al. (1994) from the [\ion{N}{ii}]/[\ion{O}{i}] ratio.

In Fig. \ref{fig:HH34_diagnostics} we also plot the values of n$_{e}$ and 
T$_{e}$ inferred from the \fe\,   lines.
As in the case of HH 111, n$_{e}$ derived from the \fe\, lines 
is higher than obtained from the \s\, lines
(n$_{e}$ $\sim$ 1 - 4.5 10$^3$ cm$^{-3}$), while the values for 
T$_{e}$ are lower
(T$_{e}$ $\sim$ 3500 - 7700 K). Such behaviour again corresponds to
the  expected structure of the shock cooling zones.  
In this jet too we find an even denser component of gas 
in the knots, as shown in Figs. \ref{fig:HH34_highdensity_Fe} and 
\ref{fig:HH34_highdensity_Ca}.
From the \fe\lam7155/\lam8617 ratio we infer values  
from  n$_{e}$ $\sim$ 4 10$^6$ cm$^{-3}$ close to the source to 
n$_{e}$ $\sim$ 10$^2$ cm$^{-3}$ in  knot L.
Similar results are obtained through \ion{Ca}{ii} diagnostics 
(Fig. \ref{fig:HH34_highdensity_Ca}).
In this case we have n$_{e}$ $\sim$ 5 - 6.3 10$^6$ cm$^{-3}$ for the inner 
knots (from A to D) and n$_{e}$ $<$ 10$^6$ cm$^{-3}$ in the outer knots 
(from E to L).
Further estimates of relevant physical quantities are discussed  in Sect. 5.

\subsection{HH 83}

\begin{figure}[!ht]
\resizebox{\hsize}{!}{\includegraphics{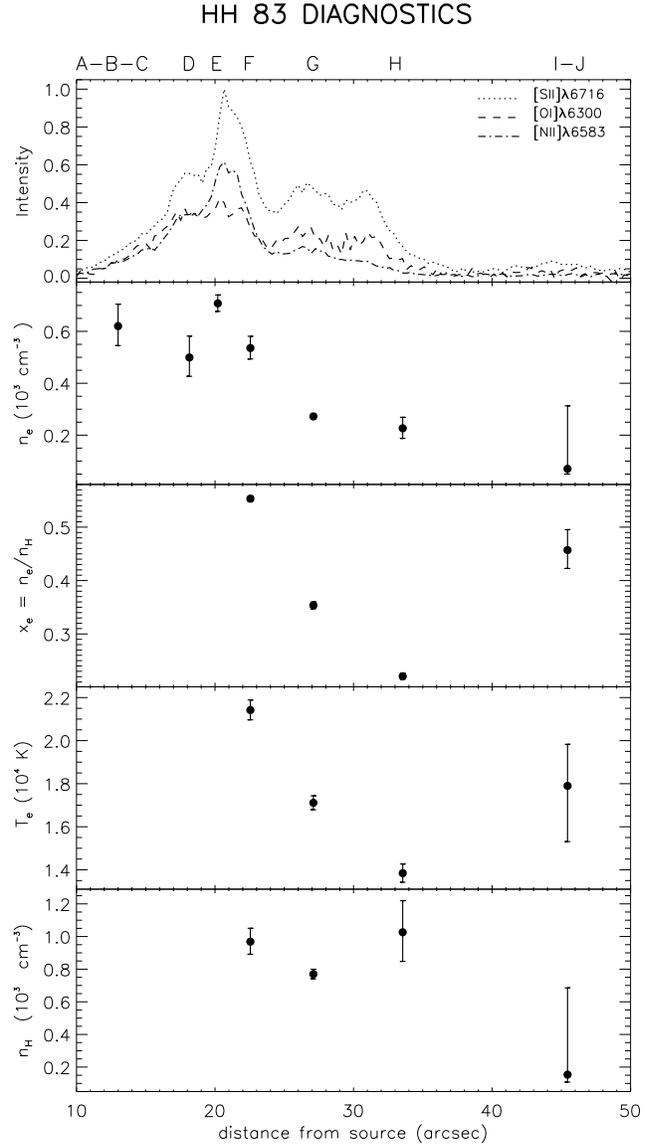}}
\caption{\label{fig:HH83_diagnostics} 
Same as Fig. \ref{fig:HH111_diagnostics}, but for the HH 83 jet.
No infrared diagnostics are available in this case (see text).
The zero point of the spatial scale is the source HH 83 IRS 
(RA(2000): 05 33 32.5, Dec(2000): -06 29 44).
}
\end{figure}

The HH 83 jet is also located in the Orion Nebula, in the L1641 molecular 
cloud (\cite{reipurth89}). 
It is powered by the IRAS source 05311-0631, 
which is detected at infrared wavelengths only from the H and K band 
(\cite{moneti95}).
The jet becomes visible in the optical and near-infrared at a distance of 
$\sim$ 10$''$ 
from the source, when it emerges from the cloud surface, detected as a 
reflection nebula (Re 17, \cite{rolph90}). 
The jet is composed of 10 blue-shifted knots (following the nomenclature
of \cite{reipurth89}) located between 10$''$ and 50$''$ from the source, and 
terminates with a bow-shock at $\sim$120$''$.
On the opposite side of the source only two faint condensations are found 
(\cite{reipurth89}).
Because of the faintness of the H$_{2}$ emission 
in the acquisition image, the slit was not well aligned for the infrared 
spectra. 
This caused the IR spectra to be degraded, and only a few, very faint \fe\, 
and H$_{2}$ lines 
were observed. Therefore 
it was not possible to extract the IR spectra for combination with the optical 
ones.
We concentrate on the optical emission from the blue lobe of the jet, 
applying only the optical diagnostics. 
From the optical \s, [\ion{O}{i}], and [\ion{N}{ii}] lines, using the BE 
technique we derive
the values of n$_{e}$,   x$_{e}$, T$_{e}$, and  n$_{H}$.
These are shown in Fig. \ref{fig:HH83_diagnostics}, and are plotted versus the 
distance from the driving source.
The electron density reaches a maximum value of $\sim$ 0.7 10$^{3}$ cm$^{-3}$ 
in knot E;
then it decreases with distance from the star down to $\sim$ 70 cm$^{-3}$ 
in knots I-J.
The ionisation fraction and the temperature could not be calculated in the 
first three knots (A-B-C, D, E) because the diagnostics did not find a 
solution for the measured line fluxes.
This is probably due to the presence of the reflection nebula, the continuum 
from which could not be properly subtracted, adding spurious flux to the jet 
emission lines.
Concerning the other knots, the ionisation fraction has a maximum value of 
$\sim$ 0.55 in  knot F, and then it 
decreases down to  $\sim$ 0.22 in  knot H. There is another local maximum in 
knot I-J, for which we found x$_{e}$ $\sim$ 0.46.
The temperature is relatively high,  T$_{e}$ $\sim$ 2.1 10$^{4}$ K in
knot F and lower in the following knots, down to 
1.4 10$^{4}$ K in knot H. Then it is higher again in knot I-J 
(T$_{e}$ $\sim$ 1.8 10$^{4}$ K).
Finally, the total density, n$_{H}$, assumes its highest value 
of $\sim$ 10$^{3}$ cm$^{-3}$ in knots F and G.
We note that a relatively higher degree of ionisation is found for
this and the following jets with respect to HH 34 and
HH 111. For an interpretation of this result, as well as 
for the determination of the mass flux in this flow, 
see the discussion in Sections 5.2 and 5.3.

\subsection{HH 73}

\begin{figure}[!ht]
\resizebox{\hsize}{!}{\includegraphics{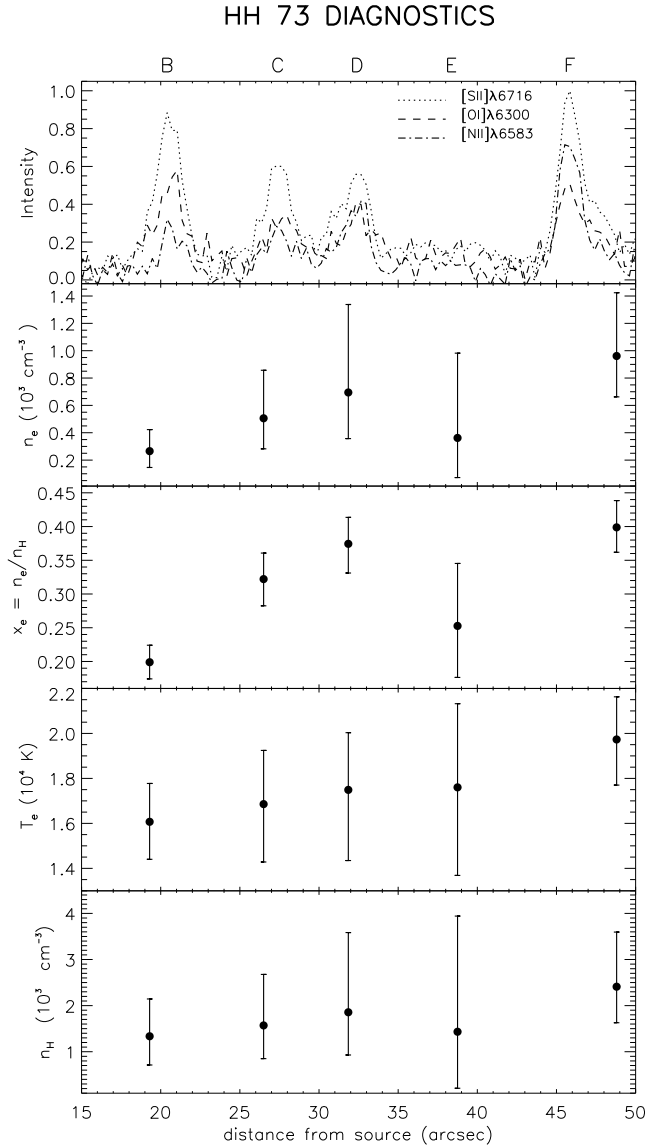}}
\caption{\label{fig:HH73_diagnostics} Same as Fig. \ref{fig:HH83_diagnostics}, 
but for the HH 73 jet.
Due to low signal-to-noise, the results are subject to large errors. 
Since the source of this jet is not known, the zero point of the spatial scale 
was set at the beginning of the first detected knot (knot A).
}
\end{figure}

The HH 73 jet, discovered by \cite{reipurth88}, is located in the Vela 
Molecular Ridge (D = 450 pc), near the small nebula Re 6 
(\cite{reipurth81}). 
It is composed of a chain of very faint optical knots, that we name from A 
to G, and extends for a total length of $\sim$ 70$''$. 
Since the source of this jet is not known, the zero point of the spatial scale 
was set at the beginning of knot A.  
We do not detect any emission in the near infrared, so we applied our 
diagnostics only to the optical emission of the brightest knots (from B to F).
The derived values of n$_{e}$, x$_{e}$, T$_{e}$, and  n$_{H}$ are shown in 
Fig. \ref{fig:HH73_diagnostics}.
Because of the low signal to noise of the line fluxes, our diagnostics is 
affected by large errors.
Nevertheless, we can define a range of variation for the physical parameters.
The values of n$_{e}$  are between $\sim$ 0.2 10$^{3}$ cm$^{-3}$
and 1.4 10$^{3}$ cm$^{-3}$, 
x$_{e}$ varies between 0.1 and 0.45, the temperature 
T$_{e}$ varies between $\sim$ 1.4 10$^{4}$ - 2.2 10$^{4}$ K, and the total 
density is in the range of $\sim$ 1-4 10$^{3}$ cm$^{-3}$.
The relatively high ionisation in this jet may be justified
following the same arguments identified for HH 83 (see Sect. 5.3).

\subsection{The HH 24 complex}

\begin{figure}[!ht]
\resizebox{\hsize}{!}{\includegraphics{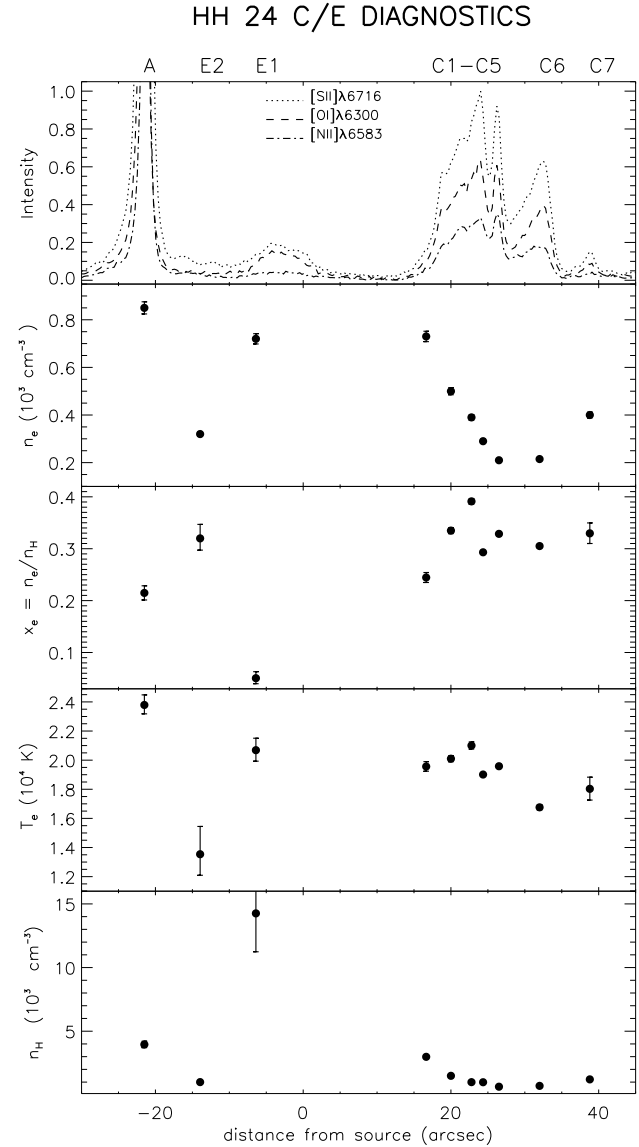}}
\caption{\label{fig:HH24C/E_diagnostics} 
Same as Fig. \ref{fig:HH83_diagnostics}, but for the HH 24 C/E jet. 
The zero point of the spatial scale is the source SSV 63, which is supposed
to be the driving source of this jet  
(RA(2000): 05 37 08.23, Dec(2000): -00 09 24.62).
}
\end{figure}

The HH 24 complex lies in the NGC 2068 nebula that includes both reflection 
and emission nebulae,
and  at least three different outflows (\cite{solf87}, \cite{mundt91},
\cite{eisloffel97}, \cite{be99}).
IR imaging by \cite{lane89} and \cite{zealey89} revealed the presence
of three optically invisible sources (SSV 63, SSV 63 W, and SSV 63 NE) which 
could be the driving sources of the observed outflows.
One of the best defined structures is the  HH24 C blue-shifted jet, aligned 
with the source  SSV 63. 
Eisl\"offel \& Mundt (1997) showed that the HH 24 C blue lobe extends for 
$\sim$1 pc, including
the ring-shaped object called HH 24 H, and terminating
with a giant bow formed by a number of HH objects (HH 20, HH 21, HH 37, HH 70).
On the opposite side there is a chain of red-shifted HH knots 
(knots of the group E, knot A, and knot M following the nomenclature of 
\cite{eisloffel97}) that seems to form the counterjet, although the structure 
of this region is still unclear.
The red lobe, in fact, deviates from the HH 24 C flow axis by 
$\sim$ 6$^{\circ}$ and there is evidence that the bright knot A may belong to 
another jet in the complex (HH 24-MMS jet according to \cite{eisloffel97}).
In our spectra we detected the innermost $\sim$80$''$ of the blue lobe 
(from knot C1 to knot C11) 
and $\sim$30$''$ of the red lobe (knots E1, E2, and A).
In the optical many lines are visible throughout all the jet length, while in 
the infrared we observed only a few faint \fe\, and \h\, lines (from the 
red-shifted knots E1 and A).
As a consequence, our diagnostics is limited to the derivation of the physical 
parameters in the region of optical emission through the BE technique 
(see Fig. \ref{fig:HH24C/E_diagnostics}).
In the blue lobe the electron density n$_e$ decreases with distance from the 
source from $\sim$ 8 10$^{2}$ cm$^{-3}$ down to $\sim$ 2 10$^{2}$ cm$^{-3}$ in 
knots C5-C6.
Then there is a local maximum in knot C7 (n$_e$ $\sim$ 4 10$^{2}$ cm$^{-3}$).
As for the red lobe, n$_e$ is equal to $\sim$ 7 10$^{2}$ cm$^{-3}$ in knot E1, 
then it decreases down to
$\sim$ 3 10$^{2}$ cm$^{-3}$ in knot E2, and finally it reaches
a value of $\sim$ 8.5 10$^{2}$ cm$^{-3}$ in knot A. 
The ionisation fraction x$_e$ varies along the jet between 0.2-0.4, with no 
definite trend.
It is lower in knot E1, where we find x$_e$ $\sim$ 0.05.
The temperature decreases with distance from the source in the blue lobe 
going from $\sim$ 2.1 10$^{4}$ K down to $\sim$ 1.7 10$^{4}$ K. 
In the red lobe, instead, we found a temperature of $\sim$ 2 10$^{4}$ K in 
knot E1, $\sim$ 1.4 10$^{4}$ K in knot E2, and a maximum of 
$\sim$ 2.4 10$^{4}$ K in knot A. 
Finally, the total density is decreasing in the blue lobe from 
$\sim$ 3 10$^{3}$ cm$^{-3}$ in knot C1
to 6 10$^{2}$ cm$^{-3}$ in knot C5, then it increases again in knots C6 and C7 
(n$_H$ $\sim$ 0.7-1.2 10$^{3}$ cm$^{-3}$).
Towards the red lobe we find n$_H$ $\sim$ 1.4 10$^{4}$ cm$^{-3}$ in
knot E1, n$_H$ $\sim$ 10$^{3}$ cm$^{-3}$ in knot E2, and 
n$_H$ $\sim$ 4 10$^{3}$ cm$^{-3}$ in knot A.
We find higher ionisation fractions and
temperatures than those derived by \cite{be99}. As
mentioned in the case of the HH 34 jet, this is due to the adoption
of a different set of abundances (see also Fig. 1). For comments on
the comparison with HH 111 and HH 34 see the discussion in Sect. 5.3.

It can be noticed that, while in the blue lobe the derived parameters have a 
definite trend along the jet,
in the red lobe the values are scattered.
We find, e.g., that the ionisation fraction and the total density in knot E1 
are, respectively, much lower and higher
than those found in the C knots.
This effect can be due to the faintness of knots E1 and E2 
(see the intensity spatial profiles in Fig. \ref{fig:HH24C/E_diagnostics})
and, as a consequence, to the low signal-to-noise ratio, which
affects, overall, the measurement of the [\ion{N}{ii}]\lam6583 line flux 
and, thus, the determination of the ionisation fraction.
It could also be intrinsic to the object, the structure of which is not 
entirely clear in this region 
(\cite{solf87}, \cite{mundt91}, \cite{eisloffel97})
(knot A may not belong to the jet and the emission from the HH 24 J jet can be 
superimposed on the knots of group E).
See further discussion about the HH 24 C/E jet in Sections 5.2 and 5.3.

Another jet in the complex is HH 24 J. 
It consists of a chain of knots and condensations, which are well aligned with 
the source SSV 63 W. 
This jet also seems to be parsec-scale, terminating with the bow HH 19 to the 
north-west and the
bow-shaped object HH 27 to the south-east (\cite{eisloffel97}).
In the optical we detect emission in the first $\sim$30$''$ of the north-west 
lobe and in 
the first $\sim$40$''$ of the south-east lobe 
(HH 24 J and HH 24 K following the nomenclature of
\cite{eisloffel97}), while no emission lines are detected in the NIR. 
Unfortunately, in the spectra of the knots,
the \s\lam\lam6716, 6731 doublet is not resolved sufficiently as to separately 
measure the line fluxes, except for two condensations located at $\sim$30$''$ 
on the north-west from the source.
Since higher resolution spectra of the HH 24 J jet are not available, we apply 
our diagnostics only to the two condensations.
For these knots we derive an electron density of $\sim$0.5 10$^{2}$
cm$^{-3}$, a ionisation fraction of $\sim$0.02 and $\sim$0.05,
a temperature of $\sim$1.5 10$^{4}$ K and $\sim$0.9 10$^{4}$ K, and a
total density of $\sim$2.5 10$^{3}$ cm$^{-3}$ and 0.9 10$^{3}$ cm$^{-3}$.

\section{Discussion}

Here we illustrate and discuss further information about the jet physics, 
derived from the basic gas parameters obtained in Sect. 4 and summarized in 
Tab. \ref{tab:physical_parameters}. 
We also compare the values of these parameters in the various objects, to 
search for common trends or differences in the sample.

\begin{table*}
\caption[]{\label{tab:physical_parameters} Physical parameters of the jets 
averaged over the brightest knots.}

\vspace{0.5cm}
    \begin{tabular}[h]{cc|cccc|cc}
      \hline \\[-5pt]
      &             &  \multicolumn{4}{c}{From O$^{0}$/S$^{+}$/N$^{+}$ lines}                  & \multicolumn{2}{c}{From Fe$^{+}$ lines}\\
Jet   & A$_{\rm V}$ & n$_e$                & x$_e$       & T$_e$        & n$_H$                &  n$_e$               & T$_e$           \\
      & (mag)       & (10$^{3}$ cm$^{-3}$) &             & (10$^{3}$ K) & (10$^{3}$ cm$^{-3}$) & (10$^{3}$ cm$^{-3}$) & (10$^{3}$ K)    \\[+5pt]
\hline \\[-5pt]
HH 111 & 2.0 - 0    & 1.0 & 0.10 & 13.0 & 11.3 & 2.6 & 7.3 \\
HH 34  & 7.1 - 1.3  & 0.8 & 0.04 & 13.8 & 16.2 & 1.8 & 5.8 \\
HH 83  &     -           & 0.5 & 0.38 & 17.5 & 0.9  &  -   & -    \\
HH 73  &      -          & 0.6 & 0.31 & 17.5 & 1.7  &  -   &   -  \\
HH 24 C/E &    -         & 0.4 & 0.32 & 19.3 & 1.3  &   -  &  -   \\[+5pt]
\hline \\
      \end{tabular}

\end{table*}

\subsection{Abundances of refractory species}

\begin{figure}[!ht]
\resizebox{\hsize}{!}{\includegraphics{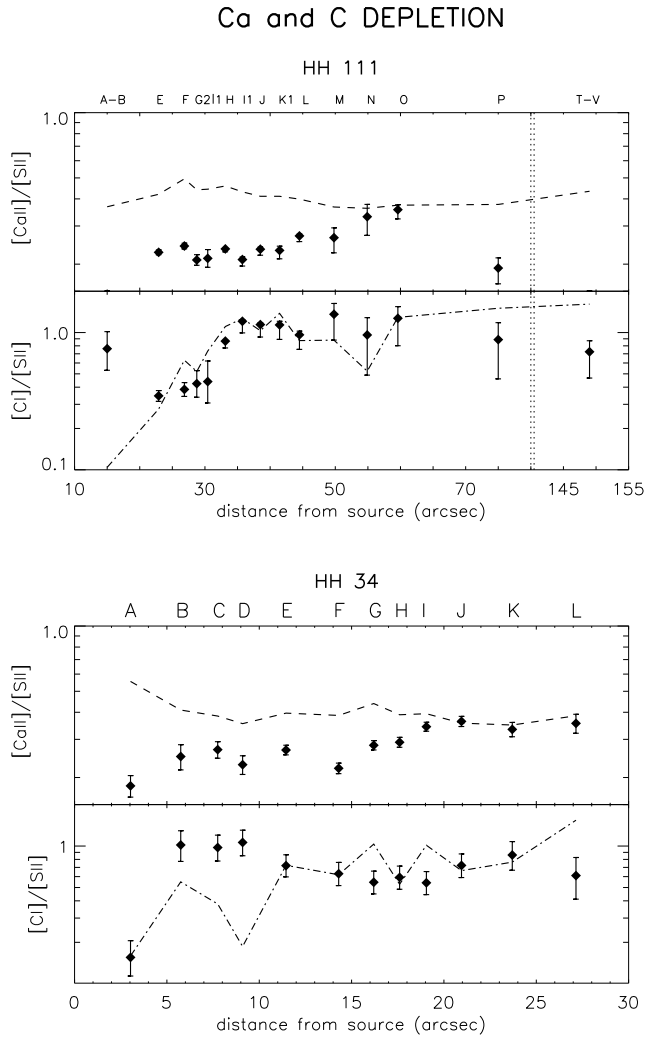}}
\caption{\label{fig:Ca_C_depletion} Comparison between observed (diamonds) 
and predicted (dashed and dash-dotted lines) 
[\ion{Ca}{ii}]$\lambda$(7290+7324)/\s$\lambda$(6716+6731) and
[\ion{C}{i}]$\lambda$(9824+9850)/[\ion{S}{ii}]$\lambda$(6716+6731) ratios 
along the HH 111 and HH 34 jets.
The theoretical ratios are computed from the derived physical parameters of 
the gas and assuming solar abundances for Ca and C. 
The [\ion{Ca}{ii}]/\s\, ratios indicate a depletion of the gas phase abundance 
of Ca of $\sim$ 50\% in the innermost knots. 
The depletion decreases with distance from the source for both HH 111 and 
HH 34. 
The good agreement between the observed and theoretical [\ion{C}{i}]/\s\, 
ratios indicates that carbon is not depleted with respect to solar values. 
Note the vertical broken lines in the HH 111 plot indicate a gap in the 
distance axis.
}
\end{figure}

\begin{figure}[!ht]
\resizebox{\hsize}{!}{\includegraphics{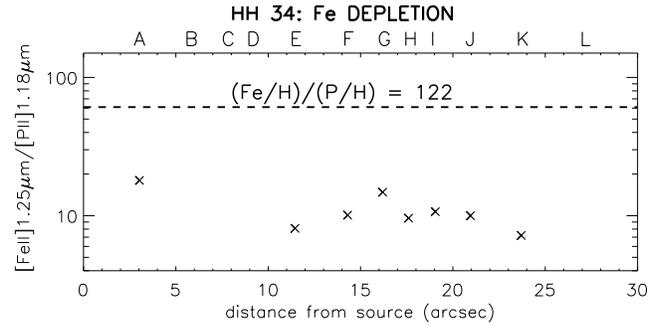}}
\caption{\label{fig:HH34_Fe_depletion} Comparison between the observed
\fe1.25\um/[\ion{P}{ii}]1.18\um\, ratios (crosses) and the expected value  
obtained assuming solar abundances for Fe and P (\cite{oliva01}). The 
position of the observed points indicates a 
depletion of $\sim$ 85 \% of the gas phase abundance of Iron. 
}
\end{figure}

\begin{table*}
\caption[]{\label{tab:depletion} Gas-phase abundance of refractory 
species with respect to the Solar Abundances determined by \cite{asplund05}.}

\vspace{0.5cm}
    \begin{tabular}[h]{c|ccc|cc}
      \hline \\[-5pt]
Species   & \multicolumn{3}{c}{[X]$_{gas}$/[X]$_{solar}$} & \multicolumn{2}{c}{Abundances: (X/H)} \\
          & HH 111      & HH 34      & Orion Cloud        & Orion Abundances$^{a}$ & Solar Abundances$^{b}$ \\[+5pt]

\hline \\[-5pt]
Ca & 0.5 - 1 & 0.3 - 1 & 0.01 & 2 10$^{-8}$    & 2.04 10$^{-6}$ \\
C  &     $\sim$ 1 &     $\sim$ 1 & 1.07 & 2.63 10$^{-4}$ & 2.45 10$^{-4}$ \\
Fe &       -      &    0.13      & 0.05 & 1.29 10$^{-6}$ & 2.82 10$^{-5}$ \\[+5pt]

\hline \\
      \end{tabular}

~$^{a}$from Esteban et al. (2004), except for the Ca abundance which was 
determined by Baldwin et al. (1991)\\
~$^{b}$from Asplund et al. (2005)

\end{table*}

The wavelength range covered by our spectra includes transitions from 
refractory species like carbon, calcium and iron.
This allows us to check for the gas phase abundances of 
these species with respect to solar values.
In the interstellar medium refractory species are often locked into dust 
grains, thus a strong depletion is expected. 
On the other hand, sputtering and photoevaporation
processes due to the passage of shock fronts can destroy all or part of
dust grains, releasing the refractory atoms into the gas cloud 
(\cite{jones00}, \cite{draine03}).
It follows that an estimate of the depletion of the gas phase abundance of 
these atoms can give important constraints on the dust structure, and on the 
efficiency of shocks in destroying grains.
The latter is related to parameters such as the shock velocity and the 
pre-shock excitation conditions.
In order to infer the elemental gas-phase abundances we compare observed and 
expected ratios between emission lines of refractory
and non-refractory species.
We select lines which are excited in the same region of the post-shocked gas, 
so that one can assume the same filling factor.

For the gas phase abundances of calcium and carbon we use the ratios
[\ion{Ca}{ii}]$\lambda$(7290+7324)/\s$\lambda$(6716+6731) and
[\ion{C}{i}]$\lambda$(9824+9850)/[\ion{S}{ii}]$\lambda$(6716+6731). 
The predicted values of these ratios are determined using the 
values of n$_{e}$,   x$_{e}$, and  T$_{e}$ inferred through the BE technique. 
Calcium is assumed to be completely ionised while the carbon 
ionisation fraction is computed considering collisional and charge exchange 
ionisation and direct and dielectronic recombination processes 
(rates from \cite{stancil98} and \cite{landini90}).
Finally, we assumed solar abundances from Asplund et al. (2005) for Ca and C.
In Fig. \ref{fig:Ca_C_depletion}  
the comparison between predicted and observed ratios along the HH 111 and 
HH 34 jets is shown.

The observed [\ion{Ca}{ii}]$\lambda$(7290+7324)/\s$\lambda$(6716+6731) ratio 
is smaller than the predicted one along both  HH 34 and HH 111. 
We interpret this discrepancy as due to depletion of Calcium atoms 
with respect to the solar value (see Tab. \ref{tab:depletion}).
Along the HH 111 jet a depletion of $\sim$ 50\% is found in the knots closest 
to the source.
Then the depletion slowly decreases with distance, down to zero in knot O. 
Another local maximum is then found in knot P.
Similarly, along HH 34 we find a maximum depletion ($\sim$70\%) in the first 
knot, which then decreases steadily until a solar gas-phase abundance is 
reached in the outermost three knots (J, K, L).
 
The trend, observed along both  HH 111 and HH 34, can be explained by the fact 
that the leading bow-shock has a much larger effective shock velocity 
(of 100-400 km s$^{-1}$) than the internal working surfaces in the beam 
(30-40 km s$^{-1}$) (\cite{raga92}).
On the other hand, the shock models predict 
substantial grain destruction only for shock speeds $>$ 100 km s$^{-1}$ 
(\cite{draine03}).
We would therefore only expect the dust to be gradually destroyed by the 
working surface as it propagate outwards.
At leading bow-shocks  however, e.g. HH 34 S, almost total destruction should 
be achieved.
Note that along the jet we measure less depletion than in the surrounding 
ambient medium (see, e.g., \cite{baldwin91}),
which is another indication that dust is partially destroyed by jet shocks.

For carbon, instead, good agreement between the observed and predicted
[\ion{C}{i}]$\lambda$(9824+9850)/[\ion{S}{ii}]$\lambda$(6716+6731) ratio
is found, at least in the brightest knots where the signal-to-noise is high.
This indicates that this species is not depleted in the Orion Cloud, 
a result that has also been found in
previous works (e.g., \cite{rubin91}, \cite{baldwin91}, \cite{peimbert93}, 
\cite{esteban04}).

In a similar way we determine the quantity of Iron atoms in gaseous form.
For Fe the comparison is complicated by the fact that the 
observed \fe\, lines come from regions of the post-shocked gas 
different from those giving rise to the optical lines. 
This is why contrasting results are found in the literature, in works that 
adopt different methods 
(see e.g. \cite{beck96}, \cite{bohm01}, \cite{nisini02}).
Since we obtained different values of n$_{e}$ and  T$_{e}$ from \fe\, and \s\, 
lines, we cannot
compare ratios between these lines, because they have probably different 
filling factors (see \cite{nisini05}).
We use, instead, the ratio \fe1.25\um/[\ion{P}{ii}]1.18\um\,. 
Phosphorus, in fact, is a non-refractory
species, and the excitation conditions of the 1.18\um\, line are similar to 
the ones of the \fe 1.25\um\, line.
The predicted \fe1.25\um/[\ion{P}{ii}]1.18\um\, is estimated to be about 
[(Fe/H)/(P/H)]/2 (\cite{oliva01}) if Iron is not depleted.
The comparison between the observed ratio in the HH 34 jet and the one 
calculated assuming solar abundances is shown in Fig. 
\ref{fig:HH34_Fe_depletion}.
From the figure we deduce that in HH 34 Fe may be depleted by 87\% with 
respect to solar ((Fe/H)$_{gas}$/(Fe/H)$_{solar}$ $\sim$ 0.13 and thus 
(Fe/H)$_{gas}$ $\sim$ 3.67 10$^{-6}$).
This depletion is lower than the one found by Esteban et al. (2004) in the 
ISM of Orion, so some atoms of Iron
have probably been unlocked by the jet shocks.
The fact that we found a different amount of depletion for Calcium and Iron
indicates that the different species follow selective patterns for their 
erosion from dust grains as expected from theory (\cite{jones00}).
Similar results have been also obtained on the HH 1 jet in Nisini et al. (2005)
indicating a common behavior of the erosion patterns in different jets.
We could not check the iron depletion along the HH 111 jet, since the 
[\ion{P}{ii}]1.18\um\, line is not visible in our spectra.

\subsection{Determination of the mass flux rate and other jet parameters}

The mass flux rate ($\dot{M}_{\rm jet}$) is a fundamental quantity
governing the jet dynamics and which enters all comparisons 
between observations and  theoretical models.
For example, in the magneto-hydro dynamic models proposed to explain jet 
formation and acceleration (see \cite{konigl00}, \cite{shu00}), 
the ratio between the rate of mass ejected into the jet ($\dot{M}_{\rm jet}$) 
and the rate of mass accreted from the disk onto the star 
($\dot{M}_{\rm acc}$) is fixed 
($\dot{M}_{\rm jet} / \dot{M}_{\rm acc} \sim 0.01 - 0.1$).
Moreover, the knowledge of $\dot{M}_{\rm jet}$ allows us to estimate other 
important dynamical quantities such as the linear ($\dot{P}_{\rm jet}$) and 
angular ($\dot{L}_{\rm jet}$) momentum fluxes carried by the jet. 
Knowing $\dot{P}_{\rm jet}$ one can check if the jet is powerful enough to 
accelerate surrounding molecular outflows, thus helping to clear the 
circumstellar environment and inject turbulence into the cloud. 
$\dot{L}_{\rm jet}$ is related instead to the jet's 
capability of removing excess angular momentum from the disk/star system, 
thus allowing the accretion of matter from the disk onto the central star 
(\cite{konigl00, woitas05}).

\begin{figure}[!ht]
\resizebox{\hsize}{!}{\includegraphics{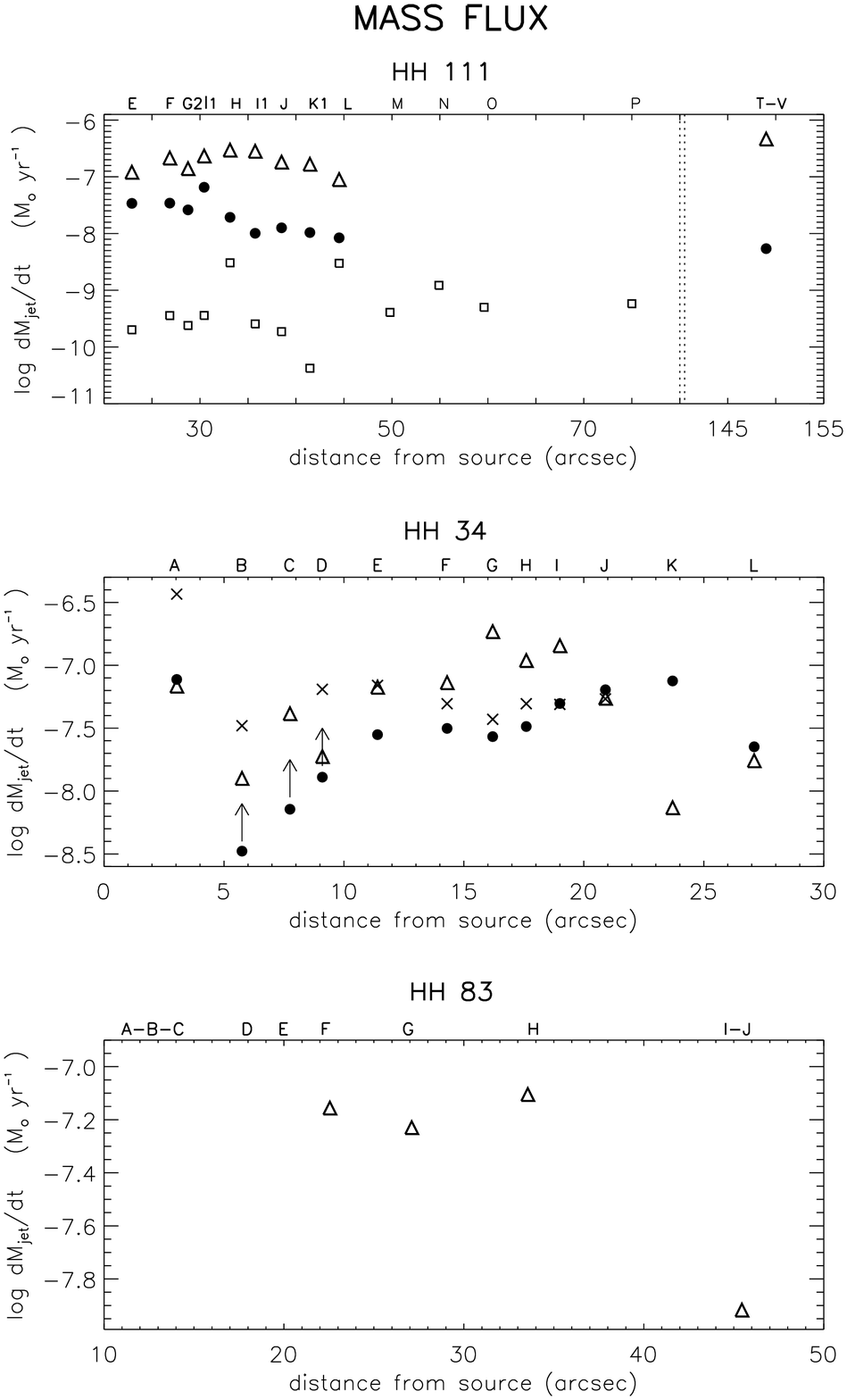}}
\caption{\label{fig:mdot}  
Mass fluxes determined for the HH 111, HH 34, and HH 83 jets. 
Triangles refer to values of  $\dot{M}_{\rm jet}$ inferred from our estimates 
of the total density, n$_{H}$, and the values of the jet radius ($r_{J}$) and 
velocity ($v_{J}$) taken from the literature (method A).
Circles  (S$^{+}$, N$^{+}$, O$^{0}$) and  crosses (\fe\,) refer to the 
estimates of $\dot{M}_{\rm jet}$ derived from line luminosities (method B). 
This method takes into account the knot beam filling.
Squares refer to the values of the mass flux transported by the molecular 
component. These values are calculated using our estimates of \h\, column 
density (method A). 
}
\end{figure}

\begin{table*}
\caption[]{\label{tab:mdot} Mass flux and linear momentum flux along the jets}

\vspace{0.5cm}
    \begin{tabular}[h]{c|ccc|cccc|c}
      \hline \\[-5pt]
Jet   & $v_{\rm jet}$$^{a}$ & $r_{\rm jet}$$^{b}$ & $n_{\rm H}$$^{c}$ & $ff$$^{d}$ & $\dot{M}_{\rm jet}$ (A)$^{e}$ & 
$\dot{M}_{\rm jet}$ (B)$^{f}$ & $\dot{P}_{\rm jet}$$^{g}$ & $\dot{P}_{\rm outflow}$$^{h}$\\
      & (\kms)      & ($''$) & (10$^{3}$ cm$^{-3}$) &   &   (M$_{\odot}$\,yr$^{-1}$) & (M$_{\odot}$\,yr$^{-1}$) 
&  (M$_{\odot}$\,yr$^{-1}$ km s$^{-1}$) &  (M$_{\odot}$\,yr$^{-1}$ km s$^{-1}$) \\[+5pt]
\hline \\[-5pt]
HH 111 & 268 & 0.25 - 1.0 & 11.3 & 0.2 & 2.2 10$^{-7}$ & 5.0 10$^{-8}$ & 1.3 10$^{-5}$ & 1.8 10$^{-5}$ \\
HH 34  & 211 & 0.15 - 0.35& 16.2 & 0.4 & 1.1 10$^{-7}$ & 3.9 10$^{-8}$ & 0.8 10$^{-5}$ & 1.2 10$^{-6}$ \\
HH 83  & 213 & 0.8             & 0.9  &  -  & 6.9 10$^{-8}$ &     -         & 1.5 10$^{-5}$ & -             \\
HH 24 C& 425 & 0.6             & 1.3  &  -  & 9.9 10$^{-8}$ &     -         & 4.2 10$^{-5}$ & -             \\[+5pt]
\hline \\
      \end{tabular}

~$^{a}$$v_{\rm jet}$ is derived from $v_{\rm t}$ and $v_{\rm r}$ estimated by: 
\cite{hartigan01} (HH 111), \cite{reipurth02} (HH 34), 
\cite{reipurth89} (HH 83) and \cite{mundt91} (HH 24 C). \\
~$^{b}$$r_{\rm jet}$ is taken to be one half of the FWHM of the \s\, intensity 
profile measured by \cite{reipurth00} (HH 111), 
\cite{reipurth02} (HH 34), \cite{mundt91} (HH 83, HH 24 C).\\
~$^{c}$$n_{\rm H}$ is the total density estimated via the BE technique (see 
Sect. 3.2). \\
~$^{d}$$ff$ is the volume filling factor estimated from the ratio between 
observed and theoretical line luminosities; this is equal to the 
ratio of $\dot{M}_{\rm jet}$ values derived from methods B and A (see text).\\
~$^{e}$$\dot{M}_{\rm jet}$ estimated through method A (see text) and averaged 
over the brightest knots; a $ff$=1 is assumed.\\
~$^{f}$$\dot{M}_{\rm jet}$ measured from \s\, and [\ion{O}{i}] line 
luminosities taking into account the beam filling (method B) and averaged 
over the brightest knots. \\
~$^{g}$$\dot{P}_{\rm jet}$ is calculated from $\dot{M}_{\rm jet}$(B) and 
$v_{\rm jet}$ ($\dot{P}_{\rm jet}$ = $\dot{M}_{\rm jet}$ $v_{\rm jet}$). \\
~$^{h}$$\dot{P}_{\rm outflow}$ is the flux of linear momentum transported by 
the molecular CO flows and measured by \cite{chernin95} (HH 34) 
and \cite{reipurth91}, \cite{cernicharo96} (HH 111). \\

\end{table*}

The determination of the physical conditions along the jet 
allows us to derive $\dot{M}_{\rm jet}$ in each knot. 
We use two different procedures, as we also did in Nisini et al. (2005) for 
the HH 1 jet.
In the first method (method A), we estimate the mass flux as
$\dot{M}_{\rm jet}$ = $\mu$\,$m_{H}$\,$n_{H}$\,$\pi$\,$r_{J}$$^2$\,$v_J$,
where $\mu$=1.24 is the mean atomic weight, $m_{H}$ the proton mass,  $n_{H}$
the hydrogen density and $r_J$ and $v_J$, respectively, the jet radius and 
velocity.
We use the total density inferred from our diagnostics and 
the values of $r_J$ and $v_J$ taken from the literature,
as our spectral resolution is too low for such an estimate  
($r_{J}$ is taken to be one half of the FWHM of the \s\, intensity profile in 
HST images).
This method is independent of the reddening estimate, but it assumes that the 
knot is uniformly filled at the density derived from the diagnostic, 
giving an upper limit to $\dot{M}_{\rm jet}$. Such an effect is 
partially compensated for by the presence of regions at even higher densities
in the beams than those traced by the \s\, lines,  which
are not taken into account in the calculations.
Alternatively (method B), $\dot{M}_{\rm jet}$ can be derived from the observed 
luminosity, $L(line)$, of 
selected optically thin lines such as \s, [\ion{O}{i}], and \fe\,, 
that is proportional to the mass of the emitting gas: 
$\dot{M}_{\rm jet}$ = $\mu$\,$m_{H}$\,($n_{H}$\,$V$)\,$v_{t}$/$l_{t}$, with 
$n_H\,V = L(line)\,\left(h\,\nu\,A_{i}\,f_{i}\,\frac{X^i}{X}\,\frac{X}{H}\right)^{-1}$,
where $V$ is the volume effectively filled by the emitting gas,
$v_{t}$ and $l_{t}$ are the tangential velocity and length of the knot, 
$A_{i}$ and $f_{i}$ the radiative rate 
and the upper level population relative to the considered transition, 
$\frac{X^i}{X}$ and $\frac{X}{H}$ are the ionisation fraction and the relative 
abundance of the considered species. 
This method is affected by uncertainties in absolute 
calibrations, extinction, and distance, but does 
implicitly take into account the volume filling factor $ff$, which, 
in practice, is simply
the ratio of $\dot{M}_{\rm jet}$ values derived from methods B and A.
In fact, the method only measures gas which is sufficiently heated
to radiate the observed lines.

The mass fluxes inferred for the HH 111, HH 34, and HH 83 jets are shown in 
Fig. \ref{fig:mdot}.
For HH 111, method A gives a mass flux approximately constant with distance 
from the source and equal to about  $\sim$ 2 10$^{-7}$ M$_{\odot}$\,yr$^{-1}$ 
($r_{J}$ are from \cite{reipurth00} and $v_{t}$ from \cite{hartigan01}).
Method B gives smaller values of $\dot{M}_{\rm jet}$, that, however, decrease 
with distance from the source.
A steady jet flow is actually expected to show a more or less constant 
$\dot{M}_{\rm jet}$ along the beam
(as derived from method A) unless significant mass is loss sideways by, e.g., 
a turbulent boundary layer or big bow shocks. 
This seems to be the case for HH 111.
A possible explanation for the observed decrease is that the jet opening angle 
increases with distance from the source to the point that  the jet diameter 
becomes larger than the slit width
from 30$''$ outwards. Thus method B determinations are affected by partial loss
of line flux.
To obtain a value of the mass flux which takes into account the beam filling, 
we use the filling factor from the knots in the first 30$''$ ($ff$ $\sim$ 0.2) 
to correct the values obtained from method A.
We obtain an average $\dot{M}_{\rm jet}$ value of 
$\sim$ 5 10$^{-8}$ M$_{\odot}$\,yr$^{-1}$. 
Finally, we estimate the mass flux transported by the molecular component 
through the inferred \h\, column density and the velocity of the molecular gas 
measured by Davis et al. (2001).
From method A we obtain $\dot{M}_{\rm jet}$(\h) 
$\sim$ 10$^{-9}$-10$^{-10}$ M$_{\odot}$\,yr$^{-1}$,
that is down two orders of magnitude with respect to the atomic component.
As explained in the previous Section, \h\, emission is thought to
arise in the C-shocks that form in the lateral wings of the bow
shocks. Although high angular resolution images in these lines are not
yet available to confirm this picture, 
the low value we find for the mass loss rate is consistent with  
\h\,  tracing only a thin shocked layer of the jet.

For HH 34 we obtain again different values from methods A and B, 
($r_{J}$ and $v_{t}$ are from \cite{reipurth02}) but in this case 
both are constant with distance, as the jet beam is always smaller than 
the slit width.
Note that the values of $\dot{M}_{\rm jet}$ derived from \fe\, luminosity are 
a lower limit since in this calculation we assumed that all Fe is in 
gaseous form.
Taking into account the beam filling, and considering only the brightest knots 
(E-J) for which we have good signal-to-noise, we infer an
average value of the mass flux of $\sim$ 3.9 10$^{-8}$ M$_{\odot}$\,yr$^{-1}$.
It can be noticed that both for HH 111 and HH 34 the values of 
$\dot{M}_{\rm jet}$ derived from method A are similar to the ones found by 
Hartigan et al. (1994) using the same technique.
The values obtained taking into account the volume filling factor (method B) 
are in both cases lower by one order of magnitude. 
The latter values are a better estimate of the mass flux transported by the 
atomic optical component.
Nevertheless, they are lower limits to the total mass flux since we are not 
considering even denser components 
nor the material that does not radiate (as shown, e.g., by the 
$\dot{M}_{\rm jet}$(\fe\,) estimates).

For HH 83 we infer the mass flux from the derived total density and the values 
of the jet radius and velocity from Mundt et al. (1991) and Reipurth (1989).
We could not estimate the mass flux from line luminosities because we do not 
have a measure of the visual 
extinction, due to the non-detection of NIR lines.
The mass flux from  method A is $\dot{M}_{\rm jet}$ $\sim$ 
6.9 10$^{-8}$ M$_{\odot}$\,yr$^{-1}$.
Lacking an estimate of $ff$ here, this is only an upper limit.

For HH 73, and HH 24 J, no literature determinations of the 
jet radius, velocity, and the angle of inclination with respect to the plane 
of the sky are available.
Thus we could not estimate the mass flux for these objects.

Finally, for HH 24 C/E we obtain raw estimates of the average mass flux from
the average jet radius and radial velocity given by Mundt et al. (1991).  
Combining our average value of the total density  
and assuming an inclination angle with respect to the plane of the sky 
of 25$^{\circ}$ (\cite{be99}) 
we find a mass flux of $\sim$ 9.9 10$^{-8}$ M$_{\odot}$\,yr$^{-1}$.

For our sample of HH jets we determine the linear momentum as 
$\dot{P}_{\rm jet}$ = $\dot{M}_{\rm jet}$ $v_{\rm jet}$,
combining  the estimated $\dot{M}_{\rm jet}$ with the average values of  
$v_J$ taken from the literature.
We find 
$\dot{P}_{\rm jet}$ $\sim$ 1.3 10$^{-5}$ M$_{\odot}$\,yr$^{-1}$\,km\,s$^{-1}$ 
for HH 111,
$\dot{P}_{\rm jet}$ $\sim$ 0.8 10$^{-5}$ M$_{\odot}$\,yr$^{-1}$ km s$^{-1}$ 
for HH 34,
$\dot{P}_{\rm jet}$ $\sim$ 1.5 10$^{-5}$ M$_{\odot}$\,yr$^{-1}$ km s$^{-1}$ 
for HH 83, and 
$\dot{P}_{\rm jet}$ $\sim$ 4.2 10$^{-5}$ M$_{\odot}$\,yr$^{-1}$ km s$^{-1}$ 
for HH 24 C.
The results are summarized in Tab. \ref{tab:mdot}.

For HH 111 and HH 34 we compare these values with those measured for the 
molecular outflows seen 
in CO lines. These were estimated by Chernin \& Masson (1995) for HH 34  and 
Reipurth \& Olberg (1991) and Cernicharo \& Reipurth (1996) for HH 111.
The comparison (see Tab. \ref{tab:mdot}) shows that the flux of 
linear  momentum carried by the HH 111 and HH 34 jets is  
higher or comparable to that of the molecular outflows.
Thus in principle the jets appear to be capable of accelerating these outflows.

In order to measure the flux of angular momentum carried away by the jet, 
we need measurements of the jet toroidal velocity, which requires 
sub-arcsecond resolution spectra (see, e.g., \cite{bacc02}, \cite{coffey04}, 
\cite{woitas05}).
To this aim one should acquire high angular and spectral  resolution
spectra with ground-based instruments equipped with adaptive optics.

\subsection{Comparison between the examined objects}

Our analysis allows us to compare physical parameters of the jets in our 
sample. 
In particular, it is interesting to investigate whether the various excitation 
conditions in our sample arise from different characteristics intrinsic to
the jets, or are set by the environment through which the jets 
propagate. 
From the analysis of the reduced spectra we can divide the examined targets 
into two classes: 
(I) jets visible both in the optical and in the near-infrared (i.e. 
HH 111 and HH 34) and 
(II) jets  that show few, very faint or no 
lines in the near-infrared (i.e. HH 83, HH 73 and HH 24 C/E).
The parameters collected in Tab. \ref{tab:physical_parameters} show that the 
jets which are not visible or
faint in the infrared are less dense than those that show NIR emission.
This is expected, since the Fe$^{+}$ levels from which infrared lines 
originate, have critical densities $>$10$^{4}$ cm$^{-3}$.
Thus, electron densities $>$10$^{3}$ cm$^{-3}$ are needed to populate them.
Jets in group (II) also show  higher ionisation fractions and temperatures. 
The fact that the ionisation fractions are higher 
in the less dense jets is consistent with the shocks
propagating in a medium of low preshock density. This
produces a higher excitation for a given
shock velocity (\cite{hartigan94}). The low electron density also
slows down the recombination process, keeping the level
of ionisation higher. An external source of UV radiation could  
also contribute to the excitation conditions. Such a source, however, remains
unidentified.

The above arguments would indicate that 
the ionizing agent is identical and has the same efficiency, 
in both groups of jets. 
On the other hand, the lower temperatures inferred for Group (I) jets
indicate that the cooling process is more efficient in the denser objects,
as expected for collisional line excitation.

We also note that the mass flux does not vary significantly 
for the two samples. If we expect 
a steap decrease of the mass flux with age, it 
seems that the sources of the two different groups have similar ages. 

HH 111 and HH 34 are two examples of giant HH flows and have similar 
characteristics such as:
(i) a well defined blue lobe consisting of a long chain of bright knots and 
an almost completely obscured red lobe;
(ii) the presence of distant bow-shocks, such as the bow T-V along the HH 111 
blue lobe at a distance of 150$''$ from the source and the HH 34 N and HH 34 S 
bows at a distance of 100$''$ from HH 34 IRS (\cite{reipurth01}));
(iii) similar velocities (see Tab. \ref{tab:mdot});
(iv) similar \s\, and \fe\, profiles decreasing with distance from the source;
(v) similar physical gas conditions (see Tab. \ref{tab:physical_parameters});
(vi) both jets have a denser component traced by [\ion{Ca}{ii}] and 
\fe\, lines;
(vii) along both jets Carbon is of solar abundance, while Calcium is depleted. 
Moreover the
depletion has a similar trend, decreasing with the distance from the source; 
(viii) similar flux of mass and linear momentum (see Tab. \ref{tab:mdot}).
On the other hand, they show important differences. 
The HH 34 jet is less extincted and both the source and the inner knots are 
visible in the optical, while the HH 111 source is deeply embedded in the 
parental cloud and the jet becomes visible only at a distance of $\sim$ 15$''$ 
from the source.
As a consequence the HH 34 jet is associated with a small molecular outflow, 
while along the HH 111 axis a powerful 
CO flow has been detected (\cite{chernin95}, \cite{cernicharo96}). 
Moreover, whereas HH 111 is a strong \h\, emitter, with a maximum 
\h\, 2.12 \um\, line flux of $\sim$ 1.4 10$^{-14}$ erg s$^{-1}$ cm$^{-2}$ from 
knots L and P, the HH 34 jet shows only a few weak \h\, lines, with a maximum 
flux of 2.5 10$^{-15}$ erg s$^{-1}$ cm$^{-2}$ along the 2.12 \um\, line from 
knot C.
This difference is probably due to the fact that the environment surrounding 
HH 34 is less dense than the one surrounding HH 111.
In fact, \h\, lines are likely to be excited in the bow wings, where
the ambient density encountered by the jet is higher than, or comparable to, 
the jet density and 
low-velocity shocks ($v \sim$ 20 km s$^{-1}$) are prevalent, 
preventing \h\, dissociation.

\section{Summary and conclusions}

The application of our combined optical/NIR diagnostics to a sample of 
protostellar jets allows us
to define their physical structure, to infer important parameters related to 
their dynamics such as their
mass and linear momentum fluxes, and to investigate  
the acceleration of associated molecular outflows, as well as the mechanism of 
dust reprocessing.
For the HH 111 and HH 34 jets, which are detected both in the optical and in 
the near infrared, we obtain the following results:

\begin{itemize}

\item[-] In HH 34 the visual extinction, $A_{\rm V}$, derived 
from \fe\, lines varies between $A_{\rm V}$$\sim$7.1 mag, at a distance of 
3$''$ from the source, 
to 1.3 mag, from 5$''$ to 30$''$ from the source. 
Along HH 111 the visual extinction is $A_{\rm V}$$\sim$2 mag between 
15$''$-30$''$ from the source  and negligible in the outer knots.

\item[-] In both jets the  electron density from the \s\, lines varies 
between  0.5-3.4 10$^{3}$ cm$^{-3}$,
while in the more compressed region traced by \fe\, lines 
we find higher electron density (n$_e$$\sim$ 1-5 10$^{3}$ cm$^{-3}$).
An even denser layer with electron densities up to 10$^{6}$ cm$^{-3}$ is 
traced by [\ion{Ca}{ii}] lines.
Such a density stratification is expected in the post-shocked gas of each 
unresolved knot.

\item[-] A temperature stratification is also found: the temperature
 derived from optical S$^{+}$, N$^{+}$ and O
lines is on average  1.3 - 1.4 10$^{4}$ K, while in the region of Iron
 emission the  temperature is always lower than 10$^{4}$ K.

\item[-] In HH 111, which is a strong \h\, emitter, we also derive the 
physical conditions of the molecular component.
These lines probably trace the lateral wings of the bow shocks
associated with the knot working
surfaces, where \h\, post-shock temperatures of 2000 - 3000 K are found.

\item[-] The ionisation fraction derived through the BE technique demonstrates 
that the gas in the jet is only
partially ionised (x$_e$ $\sim$ 0.03$\sim$0.3). The derived total
density (n$_e$/x$_e$) is about 0.1-3 10$^{4}$ cm$^{-3}$
in the region of optical emission. If we assume the same ionisation fraction 
in the more compressed regions
where \fe\, and [\ion{Ca}{ii}] lines are excited, total densities up to
10$^{7}$ cm$^{-3}$ are obtained 
in the dense layers traced by these lines.

\item[-] Estimates of the gas-phase abundance of carbon, calcium and iron are 
obtained.
We find that carbon is not depleted with respect to solar values, while 
calcium and iron are strongly depleted
(the depletion is $\sim$ 87\% for iron atoms and varies between 70\% and 0\% 
for calcium).
On the other hand, the estimated depletion is lower than the one found in the 
Orion Cloud by Baldwin et al. (1991) and Esteban et al. (2004) both for 
Calcium and Iron.  
This result demonstrates that weak shocks only partially destroy dust in the 
jet and that some grains still survive. 
In fact a velocity of 100-400 km\,s$^{-1}$, like that of a primary bow-shock,
is required to completely destroy the dust (\cite{draine03}). 
  
\item[-] Taking into account the filling factor of the gas in the knots, 
we derive mass flux rates of $\sim$ 4-5 10$^{-8}$ M$_{\odot}$\,yr$^{-1}$.
In HH 111 the flux of mass transported by the  molecular \h\, component is 
two orders of magnitude lower than the atomic one.

\item[-] The estimated flux of linear momentum is on average  
$\sim$ 10$^{-5}$ M$_{\odot}$\,yr$^{-1}$\,km\,s$^{-1}$. This is higher than,
or comparable to, the flux of momentum measured for the associated CO 
molecular outflows, and indicates that jets
can drive these molecular outflows.

\end{itemize}

For the HH 83, HH 73 and HH 24 C/E jets we could only apply optical 
diagnostics, obtaining the following results: 

\begin{itemize}
\item[-] The electron densities inferred from the \s\, doublet are very low 
(n$_e$ $<$ 10$^{3}$ cm$^{-3}$).
This explains why these HH jets are not detected in the NIR. 
In fact, the transition of \fe\, lines comes from levels which 
have a critical density higher than 10$^{4}$ cm$^{-3}$. 

\item[-] The ionisation fraction, on the other hand, is always higher in 
these jets (x$_e$ up to 0.6), consistent with the fact that for
a given shock velocity a higher ionisation is produced when the shock
propagates into a low density medium, and with the fact that recombination 
is slowed down at low electron densities. 
The latter can also be partially responsible for the higher temperatures in 
these objects (T$_e$ $\sim$ 1.4 - 3 10$^{4}$ K), because the collisional 
cooling process is less efficient at low density.
We suggest that the efficiency of the excitation mechanism  
is the same for all of the jets in our sample, and the differences in the
physical conditions of the gas are due to the fact that the HH 111 and HH 34 
jets are denser.
  
\item[-] The total density is quite low (n$_H$ $\sim$ 10$^{3}$ cm$^{-3}$). 
From these values we infer mass fluxes of $\sim$ 8 10$^{-8}$
M$_{\odot}$\,yr$^{-1}$. Note that these values are upper limits because we are 
not taking into account the filling factor of the emitting gas for these jets.

\item[-] The linear momentum fluxes are of 
$\sim$ 1-4 10$^{-5}$ M$_{\odot}$\,yr$^{-1}$\,km\,s$^{-1}$. 
\end{itemize}

The above results demonstrate the potential of our combined optical/NIR 
analysis to study the jet physical structure and represent a good basis to 
plan future higher spectral-spatial
observations with VLT-CRIRES/UVES and AO or interferometric instruments.
A higher spectral resolution, in fact, would allow us to analyse the kinematic 
structure of the jets, while with higher spatial resolution we could extend 
our analysis to the base of the jet, where it is launched and accelerated, 
testing the theoretical MHD models.

\begin{acknowledgements}
We thank Silvia Medves for letting us use her unpublished reduced spectra to
measure \s\, doublet fluxes for the HH 111, HH 83, HH 73, and HH 24 C/E jets.
We are grateful to Francesco Palla and Malcolm Walmsley for useful discussions
on elemental abundances and grain destruction processes. 
We also thank the referee for the many helpful comments to the first
version of this paper.
This work was partially supported by the European 
Community's Marie Curie Research and Training Network JETSET (Jet Simulations, 
Experiments and Theory) under contract MRTN-CT-2004-005592.
J.E. work was partially supported by the 
Deutsches Zentrum f\"ur Luft- und Raumfahrt grant 500R0009
and T.P.R.'s research by Science Foundation Ireland grant 04/BRG/P02741.
\end{acknowledgements}

\bibliographystyle{aa}

\begin{thebibliography}{}
\bibitem[{Asplund et al. 2005}]{asplund05} Asplund, M., Grevesse, N. \& Sauval, A.J. 2005, in {\it Cosmic Abundances 
as Records of Stellar Evolution and Nucleosynthesis}, ASP Conference Series, Vol. XXX (astro-ph/0410214)
\bibitem[{Bacciotti, Chiuderi \& Oliva 1995}]{bacc95}
Bacciotti, F., Chiuderi, C., Oliva, E.  1995, A\&A, 296, 185
\bibitem[{BE99}]{be99}
Bacciotti, F. \& Eisl\"offel, J. 1999, \aap, 342, 717 (BE99)
\bibitem[{Bacciotti, Eisl\"offel \& Ray, 1999}]{bacc_eisl_ray99}
Bacciotti, F., Eisl\"offel, J., Ray, T.P. 1999, A\&A, 350, 917
\bibitem[{Bacciotti 2002}]{bacciotti02} 
Bacciotti, F. 2002, RMxAC, 13, 8
\bibitem[{Bacciotti et al. 2002}]{bacc02}
Bacciotti, F., Ray, T.P., Mundt, R., Eisl\"offel, J., Solf, J.  2002, ApJ, 576, 222
\bibitem[{Bacciotti et al. 2004}]{bacc04}
Bacciotti, F., Ray, T.P., Coffey, D., Eisl\"offel, J., Woitas, J. 2004, Ap\&SS, 292, 651
\bibitem[{Baldwin et al. 1991}]{baldwin91}
Baldwin, J.A., Ferland, G.J., Martin, P.G., Corbin, M.R., Cota, S.A., Peterson, B.M., Slettebak, A. 1991, ApJ, 374, 580
\bibitem[{Bally \& Devine 1994}]{bally94}
Bally, J., Devine, D. 1994, ApJ, 428, L65
\bibitem[{Bautista \& Pradhan (1996)}]{bautista96} 
Bautista, M.A., Peng, J., \& Pradhan, A. K. 1996, ApJ, 460, 372
\bibitem[{Beck-Winchatz et al. 1996}]{beck96} 
Beck-Winchatz, B., B\"{o}hm, K.H. \& Noriega-Crespo, A. 1996,  AJ, 111, 346
\bibitem[{B\"{o}hm et al. 1980}]{bohm80}
B\"{o}hm, K.H., Brugel, E.W., Mannery, E. 1980, ApJ, 235, L137
\bibitem[{B\"{o}hm et al. 2001}]{bohm01}
B\"{o}hm, K.H. \& Matt, S. 2001, PASP, 113, 158
\bibitem[{Brugel et al. 1981}]{brugel81}
Brugel, E.W., B\"{o}hm, K.H. \& Mannery, E. 1981, ApJ Suppl., 47, 117 
\bibitem[{Cernicharo \& Reipurth 1996}]{cernicharo96}
 Cernicharo, J., Reipurth, B. 1996, ApJ, 460, 57
\bibitem[{Chernin \& Masson 1995}]{chernin95}
Chernin, L.M., Masson, C.R. 1995, ApJ, 443, 181
\bibitem[{Chidichimo 1981}]{chidichimo81} 
Chidichimo, M.C. 1981, J. Phys. B, 14, 4149
\bibitem[{Coffey et al. 2004}]{coffey04} 
Coffey, D., Bacciotti, F., Woitas, J., Ray, T.P., Eisl\"offel, J. 2004, ApJ, 604, 758
\bibitem[{Davis et al. 1996}]{davis96}
Davis, C.J., Eisl\"offel, J., Smith, M.D.  1996, ApJ, 463, 246
\bibitem[{Davis et al. 1999}]{davis99}
Davis, C.J., Smith, M.D., Eisl\"offel, J., Davies, J.K.  1999, MNRAS, 308, 539
\bibitem[{Davis et al. 2001}]{davis01}
Davis, C.J., Hodapp, K.W. \& Desroches, L. 2001, A\&A, 377, 285
\bibitem[{Devine et al. (1997)}]{devine97} 
Devine, D., Bally, J., Reipurth, B., Heathcote, S. 1997, AJ, 114, 2095D
\bibitem[{Draine 1989}]{draine89} 
Draine, B.T. 1989, in {\it Infrared Spectroscopy in Astronomy}, ESA-SP290, p. 93
\bibitem[{Draine 2003}]{draine03}
Draine, B.T. 2003, in {\it The Cold Universe: Saas-Fee Advanced Course 32}, astro-ph/0304488
\bibitem[{Eisl\"offel \& Mundt 1992}]{eisloffel92}
Eisl\"offel, J., Mundt, R. 1992, A\&A, 263, 292
\bibitem[{Eisl\"offel \& Mundt 1997}]{eisloffel97}
Eisl\"offel, J., Mundt, R. 1997, AJ, 114, 280
\bibitem[{Eisl\"offel, Smith and Davis 2000}]{eislsmithdavis00}
Eisl\"offel, J., Smith, M.D., Davis, C.J. 2000, A\&A, 359, 1147
\bibitem[{Eisl\"offel et al. 2000}]{eisloffel00}
Eisl\"offel, J., Mundt, R., Ray, T.P., Rodr\'{\i}guez, L.F.  2000, Protostars and Planets IV, 
Tucson: University of Arizona Press, eds Mannings, V., Boss, A.P., Russell, S.S., p. 815
\bibitem[{Esteban et al. 2004}]{esteban04}
Esteban, C., Peimbert, M., García-Rojas, J., Ruiz, M. T., Peimbert, A., Rodríguez, M. 2004, MNRAS, 355, 229
\bibitem[{Flower et al. 2003}]{flower03}
Flower, D.R., Le Bourlot, J., Pineau des Forêts, G., Cabrit, S. 2003, Ap\&SS, 287, 183
\bibitem[{Giannini et al. 2004}]{giannini04}
Giannini, T., McCoey, C., Caratti o Garatti, A., Nisini B., Lorenzetti
D. \& Flower, D.R. 2004, A\&A, 419, 999
\bibitem[{Gredel et al. 1993}]{gredel93}
Gredel, R., Reipurth, B. 1993, ApJ, 407, L29
\bibitem[{Grevesse \& Sauval 1998}]{grevesse98} Grevesse, N. \& Sauval, A.J. 1998, Space Sci. Rev. 85, 161
\bibitem[{Gueth et al. 1999}]{gueth99}
Gueth, F., Guilloteau, S.  1999, A\&A, 343, 571
\bibitem[{Jones 2000}]{jones00}
Jones, A.P. 2000, \jgr, 105, 10257
\bibitem[{Hartigan et al. 1987}]{hartigan87}
Hartigan, P., Raymond, J., Hartmann, L. 1987, ApJ, 316, 323
\bibitem[{Hartigan et al. 1994}]{hartigan94}
Hartigan, P., Morse, J.A. \& Raymond, J. 1994, ApJ, 444, 943
\bibitem[{Hartigan et al. 2001}]{hartigan01}
Hartigan, P., Morse, J.A., Reipurth, B., Heathcote, S., Bally, J. 2001, ApJ, 559, 157  
\bibitem[{Hartigan 2003}]{hartigan03}
Hartigan, P. 2003, Ap\&SS, 287, 111
\bibitem[{Hartigan et al. 2004}]{hartigan04}
Hartigan, P., Edwards, S. \& Pierson, R. 2004, ApJ, 609, 261
\bibitem[{Hillenbrand \& White 2004}]{hillenbrand04}
Hillenbrand, L.A., White, R.J. 2004, ApJ, 604, 741
\bibitem[{Hirth et al. 1997}]{hirth97}
Hirth, G.A., Mundt, R., Solf, J. 1997, A\&AS, 126, 437
\bibitem[{Hollenbach 1989}]{hollenbach89}
Hollenbach, D., McKee, C.F. 1989, ApJ, 342, 306
\bibitem[{Hollenbach 1997}]{hollenbach97}
Hollenbach, D. 1997, IAUS, 182, 181
\bibitem[{K\"onigl \& Pudritz 2000}]{konigl00}
K\"onigl, A., Pudritz, R.E. 2000, Protostar and Planets IV, 
Tucson: University of Arizona Press, eds Mannings, V., Boss, A.P., Russell, S.S., p. 759
\bibitem[{Landini \& Monsignori Fossi 1990}]{landini90} 
Landini, M. \& Monsignori Fossi, B.C. 1990, A\&AS, 82, 229
\bibitem[{Lane (1989)}]{lane89}
Lane, A.P., 1989, in ESO Workshop on {\it Low Mass Star Formation and Pre-Main Sequence Objects}, Ed. B. Reipurth (ESO, Garching), p. 331
\bibitem[{Medves, Bacciotti \& Eisl\"{o}ffel, in prep.}]{medves03}
Medves, S., Bacciotti, F., Eisl\"{o}ffel, J. 2006, in preparation
\bibitem[{Mendoza 1983}]{mendoza83} 
Mendoza, C. 1983, in Planetary Nebulae, ed. D. R. Flower (Dordrecht: Reidel), IAU Symp., 103, 143
\bibitem[{Moneti and Reipurth 1995}]{moneti95}
Moneti, A., Reipurth, B. 1995, A\&A, 301, 721
\bibitem[{Morse et al. 1992}]{morse92}
Morse, J.A.,  Hartigan, P., Cecil, G., Raymond, J., Heathcote, S.
1992, ApJ, 399, 231
\bibitem[{Morse et al. 1993}]{morse93}
Morse, J.A., Heathcote, S., Hartigan, P., Cecil, G. 1993, AJ, 106, 1139
\bibitem[{Mundt et al. 1991}]{mundt91}
Mundt, R., Ray, T.P., Raga, A.C. 1991, A\&A, 252, 740
\bibitem[{Mouri and Taniguchi 2000}]{mouri00} 
Mouri, H. \& Taniguchi, Y. 2000, ApJL, 534, L63
\bibitem[{Nisini et al. 2002}]{nisini02} 
Nisini, B., Caratti o Garatti, A., Giannini, T., \& Lorenzetti, D. 2002, A\&A, 393, 1035
\bibitem[{Nisini, Antoniucci, Giannini \& Lorenzetti 2005}]{nisini_ant05}
Nisini, B., Antoniucci, S., Giannini, T., Lorenzetti, D. 2005, A\&A, 429, 543
\bibitem[{Nisini et al. 2005}]{nisini05} Nisini, B., Bacciotti, F., Giannini, T., Massi F., 
Eisl\"{o}ffel, J., Podio, L., and Ray, T.P. 2005, A\&A, in press
\bibitem[{Noriega-Crespo et al. 1993}]{noriega93}
Noriega-Crespo, A., Garnavich, P.M., \& Raga, A.C. 1993, AJ, 106, 1133
\bibitem[{Nussbaumer \& Storey (1988)}]{nussbaumer88} 
Nussbaumer, H. \& Storey, P.J. 1988, A\&A, 193, 327
\bibitem[{Oliva et al. 2001}]{oliva01}
Oliva, E., Marconi, A., Maiolino, R. et al. 2001, A\&A, 369, L50
\bibitem[{Osterbrock 1994}]{osterbrock94}
Osterbrock D.E., {\it Astrophysics of Gaseous Nebulae and Active Galactic Nuclei}, Mill Valley: University Science Book (1994)
\bibitem[{Pesenti et al. 2003}]{pesenti03}
Pesenti, N., Dougados, C., Cabrit, S., O'Brien, D., Garcia, P., Ferreira, J. 2003, A\&A, 410, 155
\bibitem[{Peimbert et al. 1993}]{peimbert93}
Peimbert, M., Torres-Peimbert, S., Dufour, R.J.  1993, ApJ, 418, 760
\bibitem[{Quinet et al. 1996}]{quinet96}
Quinet, P., Le Dourneuf, M., Zeippen, C.J. 1996, A\&AS, 120, 361
\bibitem[{Raga \& B\"ohm (1986)}]{raga86}
Raga, A.C., B\"ohm, K.H. 1986, ApJ, 308, 829
\bibitem[{Raga \& Kofman 1992}]{raga92}
Raga, A.C., Kofman, L. 1992, ApJ, 386, 222
\bibitem[{Raga et al. 2002}]{raga02}
Raga, A. C., Noriega-Crespo, A., Reipurth, B., Garnavich, P. M.,
Heathcote, S., B\"ohm, K. H., Curiel, S. 2002, ApJ, 565, L29
\bibitem[{Ray et al. 1996}]{ray96}
Ray, T.P., Mundt, R., Dyson, J.E., Falle, S.A.E.G., Raga, A.C. 1996, ApJ, 468, L103
\bibitem[{Ray et al. 2003}]{ray03}
Ray, T.P., Bacciotti, F. 2003, RMxAC, 15, 106
\bibitem[{Reipurth 1981}]{reipurth81}
Reipurth, B. 1981, A\&A, 44, 379
\bibitem[{Reipurth et al. 1986}]{reipurth86}
Reipurth, B., Bally, J., Graham, J.A., Lane, A.P., Zealey, W.J. 1986, A\&A, 164, 51 
\bibitem[{Reipurth \& Graham (1988)}]{reipurth88}
Reipurth, B., Graham, J.A. 1988, A\&A, 202, 219 
\bibitem[{Reipurth 1989}]{reipurth89}
Reipurth, B. 1989, A\&A, 220, 249
\bibitem[{Reipurth \& Olberg 1991}]{reipurth91}
Reipurth, B., Olberg, M. 1991, A\&A, 246, 535
\bibitem[{Reipurth et al. 1997}]{reipurth97}
Reipurth, B., Hartigan, P., Heathcote, S., Morse, J.A., Bally, J. 1997, AJ, 114, 757
\bibitem[{Reipurth et al. 2000}]{reipurth00}
Reipurth, B., Yu, K.C., Heathcote, S.,  Bally, J. \& Rodriguez, L.F. 2000, AJ, 120, 1449 
\bibitem[{Reipurth \& Bally 2001}]{reipurth01}
Reipurth, B., Bally, J. 2001, ARA\&A, 39, 403
\bibitem[{Reipurth et al. 2002}]{reipurth02}
Reipurth, B., Heathcote, S., Morse, J., Hartigan, P., Bally, J. 2002, AJ, 123, 362
\bibitem[{Rieke \& Lebofsky (1985)}]{rieke85} Rieke, G. H., \& Lebofsky, M. J. 1985, ApJ, 288, 618
\bibitem[{Rolph et al. 1990}]{rolph90}
Rolph, C.D., Scarrott, S.M., Wolstencroft, R.D. 1990, MNRAS, 242, 109
\bibitem[{Rubin et al. 1991}]{rubin91}
Rubin, R.H., Simpson, J.P., Haas, M.R., Erickson, E.F. 1991, ApJ, 374, 564
\bibitem[{Shang et al. 2002}]{shang02}
Shang, H., Glassgold, A.E., Shu, F.H., Lizano, S. 2002, ApJ, 564, 853
\bibitem[{Shu et al. 2000}]{shu00}
Shu, F.H., Najita, J.R., Shang, H., Li, Z-Y  2000, Protostar and Planets IV, 
Tucson: University of Arizona Press, eds Mannings, V., Boss, A.P., Russell, S.S., p. 789
\bibitem[{Smith \& Hartigan 2006}]{smith06}
Smith, N., Hartigan, P. 2006, ApJ, 638, 1045
\bibitem[{Solf 1987}]{solf87}
Solf, J. 1987,  A\&A, 184, 322
\bibitem[{Stancil et al. 1998}]{stancil98} 
Stancil, P.C., Havener, C.C., Krstic, P.S. et al. 1998, ApJ, 502, 1006
\bibitem[{Stanke et al. 1998}]{stanke98}
Stanke, T., McCaughrean, M.J., Zinnecker, H. 1998, A\&A, 332, 307
\bibitem[{Stapelfeldt et al. 1991}]{stapelfeldt91}
Stapelfeldt, K.R., Scoville, N.Z., Beichman, C.A., Hester, J.J., Gautier III, T.N. 1991, ApJ, 371, 226
\bibitem[{Wilson \& Rood 1994}]{wilson94}
Wilson, T.L., Rood, R.T. 1994, ARA\&A, 32, 191
\bibitem[{Woitas et al. 2005}]{woitas05}
Woitas, J., Bacciotti, F., Ray, T.P., Marconi, A., Coffey, D., Eislöffel, J.  2005, A\&A, 432, 149
\bibitem[{Zealey et al. (1989)}]{zealey89}
Zealey, W.J., Mundt, R., Ray, T.P., Sandell, G., Geballe, T., Taylor, K.N.R., Williams, P.M., Zinnecker, H. 1989, PASAu, 8, 62
\end{thebibliography}

\end{document}